\documentclass{article}
\usepackage{amsmath,amsfonts,amssymb,latexsym} 
\usepackage{graphicx}
\usepackage[]{epsfig}
\usepackage{bm}
\usepackage{stmaryrd}
\newcommand{\beq}{\begin{equation}}
\newcommand{\eeq}{\end{equation}}
\newcommand{\beqd}{\begin{displaymath}}
\newcommand{\eeqd}{\end{displaymath}}
\newcommand{\beqa}{\begin{eqnarray}}
\newcommand{\eeqa}{\end{eqnarray}}

\newcommand{\comment}[1]{}

\newcommand{\grad}{\nabla}

\begin{document}

\title{Field Theory of Fluctuations in Glasses}

\author{Silvio Franz\\ {\small Laboratoire de Physique Th\'eorique et Mod\`eles
    Statistiques,} \\ {\small CNRS et Universit\'e Paris-Sud 11,
    UMR8626, B\^at. 100, 91405 Orsay Cedex, France}\\
\\
Giorgio Parisi, Federico Ricci-Tersenghi, Tommaso Rizzo\\
{\small Dipartimento di Fisica, INFN -- Sezione di Roma I, IPFC-CNR --
  UOS Roma}\\ {\small Sapienza Universit\`a di Roma, P.le Aldo Moro 2,
  I-00185 Roma, Italy}}

\maketitle

{\bf Abstract}\\
We develop a field-theoretical description of dynamical
heterogeneities and fluctuations in supercooled liquids close to the
(avoided) MCT singularity.  Using quasi-equilibrium arguments we
eliminate time from the description and we completely characterize
fluctuations in the beta regime. We identify different sources of
fluctuations and show that the most relevant ones are associated to
variations of ``self-induced disorder'' in the initial condition of
the dynamics. It follows that heterogeneites can be describes through
a cubic field theory with an effective random field term. The
phenomenon of perturbative dimensional reduction ensues, well known in random
field problems, which implies an upper critical dimension of the
theory equal to 8. We apply our theory to finite size scaling for
mean-field systems and we test its prediction against numerical
simulations.

\section{Introduction}
\label{intro}

The heterogeneous character of glassy dynamics has been object of
extensive study in the last decade \cite{het}. Experiments,
simulations and theory converge to a description of supercooled
liquids where, on approaching the glass transition, relaxation
requires cooperative motions on high mobility regions of increasing
size and life time.  An important theoretical step in the
understanding of dynamical hetrogeneities has consisted in realize
that the current theory of glassy dynamics, the Mode Coupling Theory
(MCT) \cite{MCT}, predicts a growing dynamical length as the Mode
Coupling critical point is approached. This was first 
noted
 in the
context of disordered mean field systems where MCT is exact
\cite{FPchi4}, and later confirmed with diagrammatic approaches to the
dynamics of liquids \cite{BB0}.  In the resulting picture, the
dynamical heterogeneities are captured by a time dependent four point
correlation function, whose associated dynamical length diverges at
the Mode Coupling critical point. As it is well known, this
divergence, which is genuine in mean-field, is in real systems an
artefactual consequence of MCT that neglects activated processes. The
divergence is cut-off as the MCT dominated regime at high temperature
crosses over to the barrier dominated regime at low temperature. With
this caveat, the MCT prediction of a pseudo-critical growth of
dynamical correlations has been largely confirmed in numerical
simulations \cite{numerical} and experiments \cite{pp}.  However,
corrections to MCT are at work as soon as the mean-field approximation
is not exact.  Two kinds of corrections to MCT can be expected: those
due to critical fluctuations which are not well described by
mean-field theory, and those due to barrier jumping processes.
Clarification of both kind of fluctuations is necessary to have an
accomplished theory of glassy dynamics. Unfortunately both kind of
phenomena are poorly understood.

The goal of this paper is to present an in-depth analysis of
perturbation theory around MCT to study critical fluctuations. In
doing that we will neglect barrier jumping which is intrinsically of
non-perturbative nature.

The Mode Coupling (MC) approximation describes an ergodicity breaking
transition where a system prepared in an equilibrium initial condition
remains confined in its vicinity. Correspondingly, two point connected
correlation functions develop an infinitely long plateau.  This
ergodicity breaking can be interpreted in the broader perspective of
Random First Order Theory \cite{RFOT}. 
This theory predicts that
 within the approximations in
which MCT is valid, at the dynamical transition the space of
equilibrium configuration is partitioned in in an exponentially large
number of metastable states.  Several aspects of dynamical freezing
can then be conveniently studied through equilibrium techniques,
introducing appropriate constraints in the Boltzmann-Gibbs measure
\cite{FPpot}.  The free-energy as a function of the constraints
provides a purely static field theoretical description of the MC
ergodicity breaking transition. This description has indeed been
crucial to the first theoretical recognition of the growth of a
dynamical susceptibility at a MC transition \cite{FPchi4}.  In this
paper, we exploit this constrained equilibrium technique to devise a
theory of critical fluctuations. The various dynamical
characterization of fluctuations will be expressed in {\it
  reparametrization invariant} form eliminating the time dependence in
favor of a dependence on the average value of the (two point)
correlation function itself.\footnote{The terminology is mutated from
  asymptotic aging theory where time dependence is expressed through
  dependence on average correlations \cite{CuKu}.}  This perspective
allows enormous simplification with respect to the dynamical
perturbation theory \cite{BBBKMR} which at present is limited to the
gaussian approximation.  Previous studies have stressed the importance
of emerging reparametrization invariance at large times as a soft mode
of fluctuations in \cite{leticia} glassy dynamics.  Our approach will
allow us to give a universal description of these modes in the beta
regime where dynamical correlation functions are close to their
plateau value.

The main thesis of this paper is that reparametrization invariant
fluctuations for temperatures close to the mode coupling critical
temperature $T_d$ and values of the correlations close to the plateau
value can be described in terms of a field theory of the kind
\begin{eqnarray}
H[\phi|\delta \epsilon(x)]=\int dx\; \frac 1 2 (\grad
\phi(x))^2+(\epsilon +\delta\epsilon(x))\phi(x) +g\phi^3(x)
\label{effham}
\end{eqnarray}
where $\phi(x)$ is a local fluctuation of the overlap away from the
plateau value, $\epsilon=T-T_d$ is the deviation from the critical
temperature, $g$ is a coupling constant and $\delta \epsilon (x)$ is
an effective random temperature term, distributed with gaussian
statistics and delta correlated in space.  The effective Hamiltonian
(\ref{effham}) coincides with the one that describes the spinodal
point of the Random Field Ising model (RFIM) \cite{RFIM}.  We find
 that both problems are perturbatively in the same universality
class.  The random temperature term is the ultimate consequence of
dynamic heterogeneity and is a formal expression of ``self-induced
disorder'' sometimes advocated to describe structural glasses. The
role of this term is crucial. Random field models are well studied
systems. It is well known that the random field changes the singular
behavior of the theory.  In particular in perturbation theory one
finds the phenomenon of ``dimensional reduction'' which states that
the singularities of the random model in dimension $D$ are identical
to the ones in absence of disorder in dimension $D-2$. It follows that
the upper critical dimension above which fluctuations can be expected
to have a Gaussian nature is found to be eight rather then six as it
could expected from a pure $\phi^3$ theory. It remains to find out if
the Random Field Ising Model has a relevance for glassy dynamics
beyond perturbation theory in the barrier dominated regime.

The rest of the paper is organized as follows: in section
\ref{sec:corr} we analyze the sources of fluctuations in the systems
and we define correlation functions sensitive to them. In section
\ref{due} we discuss constrained measures. We explain their use in the
computation of correlation functions and how to obtain them from
replica field theory. In section \ref{replica-action} we analyze the
replica field theory close to the MC critical temperature and study
the quadratic fluctuations.  In section \ref{pert} we analyze deeply
perturbation theory and we derive the effective field theory
(\ref{effham}).  Section \ref{f-s-s} is devoted to finite size scaling
in mean-field systems. The results of this last analysis are compared
with numerical simulations in section \ref{sec:simu}.  Finally we
expose some concluding remarks in \ref{concl}.

A partial account of the theory and simulations exposed in this paper
has been given in \cite{FRRP1}.

\section{Measures of Fluctuations}
\label{sec:corr}

The theory exposed in this paper will be largely independent on the
choice of systems. The main hypothesis we will make is that in some
approximation a MC transition is present and we will study the generic
behavior of fluctuations around it. Our theory apply equally well to
describe critical fluctuations around the avoided MC transition in
liquids as well as finite size scaling around MCT in mean-field spin
models where the transition is sharp in the thermodynamic limit. With
the former application in mind, in the following we will use the
language of field theory. In our formulae finite size scaling in
Mean-Field models can be obtained simply replacing all space
integrations by an overall volume factor $N$.

For notational convenience we will represent the systems in terms of
spin variables fixed in space $S_i=\pm 1$, $i=1,...,N$. With this
notation we can equally well describe genuine spin systems like spin
glasses, but also liquid systems in a lattice gas representation where
we divide the volume in small cells and use the spin -taking the two
values $\pm 1$- to represent the occupancy of the
cells.\footnote{Having in mind a monodisperse systems occupying a $D$
  dimensional box of linear size $L$, we can divide the volume in
  $N=(L/a)^D$ cells of linear size $a$ of the order of a fraction of
  the particle diameter. We then assign to each cell $i$ the variable
  $S_i$ which takes the value 1 if the center of a particle lies in
  the box and the value -1 otherwise.}  We will use as order parameter
of freezing the correlation function, or overlap, among spin
configurations. Given two configurations of the system $S$ and $S'$ we
can define the local value of the overlap coarse-grained over some
volumes $v$ containing a large number of spins $|v|\gg 1$,
$q_x(S,S')=|v|^{-1} \sum_{i\in v_x} S_i S_i'$. Different notions of
correlations among configurations {\it e.g.} the one used in \cite{PL}
lead to the same results, modulo a redefinition of the non-universal
parameters appearing in (\ref{effham}).  If we denote by $S(t)$ the
configuration of the system at time $t$, the time dependent
correlation function can be written as $C(t,0)=\frac 1 V \int_V dx
q_x(S(0),S(t))$. The objects of our analysis will be the fluctuations
in the global quantity $C(t,0)$ and the local quantities
$q_x(S(0),S(t))$ as they can be studied through 4-point or higher
order correlation functions.

We would like to separate the contributions of different source of
fluctuations of $C (t, 0)$. For structural glasses we would like to
distinguish fluctuations among different trajectories that start from
the same initial configuration from fluctuations due to changes in the
initial condition itself.  Recent numerical studies in supercooled
liquids have emphasized the importance of this separation to study the
influence of the structure in the development of dynamical
heterogeneities \cite{harrowell}.  For systems with quenched disorder,
like {\it e.g.}  spin glasses, one has a third source of fluctuations
in the choice of the quenched interactions. In the following we assume
without loss of generality the presence of some quenched disorder. If
there is no disorder the respective averages are immaterial.  We
denote by $\langle \cdot \rangle$ the average over trajectories that
start from the same initial condition. This was called
iso-configurational average in \cite{harrowell}.  The
iso-configurational average can be the average over the initial
velocities in the case of Newtonian dynamics or the average over thermal
noise along the trajectories in the case of stochastic heat bath
dynamics.  The initial condition is denoted by $S(0)=S^0$ and will
always be chosen as an equilibrium configuration in this paper.  The
corresponding average will be denoted by $\llbracket \cdot
\rrbracket$.  Finally averages over quenched disorder will be denoted
by $\mathbb{E}(\cdot)$.  A widely used measure of dynamical
correlations is the 4-point correlation \cite{DAS} $\chi_4(t)=N
\mathbb{E}\llbracket \langle C(t)^2\rangle\rrbracket
-\left(\mathbb{E}\llbracket \langle C(t)\rangle\rrbracket \right)^2$.
In order to quantify the contribution of each source of noise to this
function we use a decomposition of $\chi_4$ in three different terms
$\chi_4=\chi_{th}+\chi_{het}+\chi_{dis}$ defined as
\cite{Berthier-Jack}
\begin{eqnarray}
&&\frac1N \chi_{th}(t)= \mathbb{E} (\llbracket \langle C(t,0)^2
  \rangle \rrbracket )-\mathbb{E} (\llbracket \langle C(t,0) \rangle^2
  \rrbracket ) 
\nonumber \\
\;
\nonumber \\
&& \frac1N \chi_{het}(t)= \mathbb{E}({ \llbracket }\langle C(t,0)
\rangle^2 \rrbracket )-\mathbb{E} (\llbracket \langle C(t,0) \rangle
\rrbracket ^2) 
\nonumber \\
\;
\nonumber \\
&&\frac1N \chi_{dis}(t)= \mathbb{E} (\llbracket \langle C(t,0) \rangle
\rrbracket ^2)-\mathbb{E}(\llbracket \langle C(t,0) \rangle \rrbracket
)^2. 
\end{eqnarray}
These susceptibilities are the space integral of correlation functions
that we will denote respectively $G_{th}(x,t)$, $G_{het}(x,t)$ and
$G_{dis}(x,t)$. For example $G_{het}(x,t)$ can be expressed as:
\begin{eqnarray}
&&G_{het}(x,t)= \mathbb{E}(  \llbracket \langle q_x(S(0),S(t)) \rangle
  \langle q_0(S(0),S(t)) \rangle \rrbracket - \llbracket \langle
  q_x(S(0),S(t)) \rangle\rrbracket \llbracket  \langle q_0(S(0),S(t))
  \rangle \rrbracket ).
\label{4point}
\end{eqnarray}
In the case of liquids where quenched disorder is absent one has
$\chi_{dis}=0$ and 
\begin{eqnarray}
&&\frac1N \chi_{th}(t)= \llbracket \langle C(t,0)^2 \rangle \rrbracket -\llbracket \langle C(t,0) \rangle^2 \rrbracket 
\nonumber \\
\;
\nonumber \\
&&\frac1N \chi_{het}(t)= { \llbracket }\langle C(t,0) \rangle^2 \rrbracket -\llbracket \langle C(t,0) \rangle \rrbracket ^2.
\end{eqnarray}

In the following we will analyze the behavior of these three
characterizations of fluctuations and predict their behavior for times
such that the average correlation function $C_{av}(t)=\mathbb{E}
(\llbracket \langle C(t,0) \rangle \rrbracket )$ is close to the
plateau value $C_p$. As we will see in next section this can be
achieved through quasi-equilibrium techniques at the price of
eliminating time from the description. In the aforementioned time
regime $C_{av}(t)$ is a decreasing function of time.  We can express
time dependence through the dependence on $C_{av}(t)$ itself. For any
time dependent quantity $O(t)$ we write $O(C)=O(t)|_{C_{av}(t)=C}$.
All the time dependence is condensed in the dependence of $C_{av}(t)$
on time that we will leave unspecified.

\begin{figure}[ht]
\begin{center}
\includegraphics[width= 0.7 \textwidth]{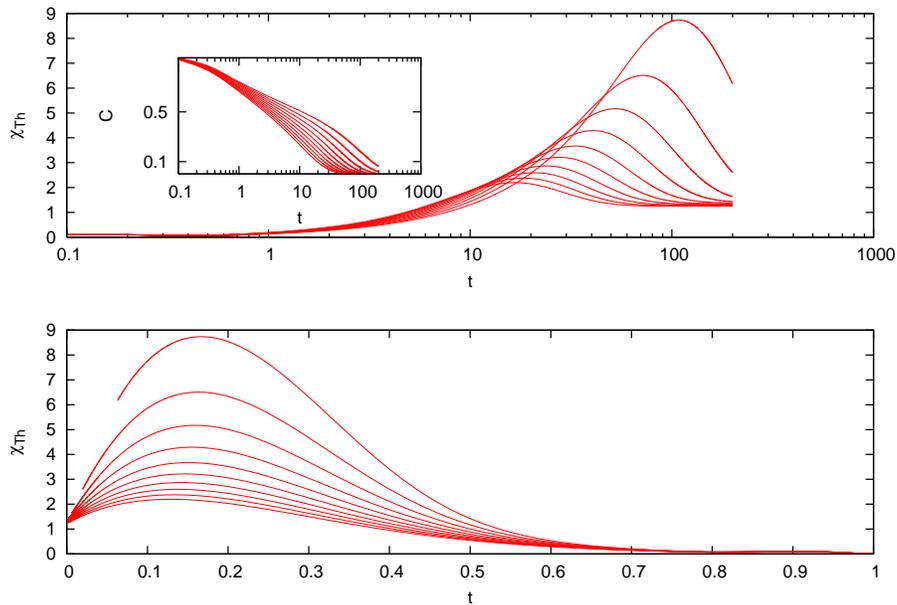}
\caption{The behavior of $\chi_{th}$ for the spherical p-spin model
  with $p=3$. The curves are at temperatures $T=0.665, 0.68, 0.695,
  0.71, 0.725, 0.74, 0.77, 0.785, 0.8$ the critical temperature being
  $T_d=\sqrt{3/8}=0.612$. Upper panel $\chi_{th}$ as a function of
  time. Inset: $C$ as a function of time. Lower panel $\chi_{th}$ as a
  function of $C$. }
\label{champ}
\end{center}
\end{figure}

\section{Quasi-equilibrium in Dynamics and 
Constrained Boltzmann-Gibbs measures.}
\label{due}

In this section we discuss how to obtain information about equilibrium
dynamics through the use of constrained equilibrium measures.  This
possibility relies in the phenomenon of time scale separation observed
in glassy dynamics, where one can separate the degrees of freedom in
fast and slow ones.

The dynamics of liquids close to the glass transition can be described
as a slow process where the system passes from one metastable state to
another. Time scale separation tells us that approximate equilibrium
establishes in a given metastable state before a new state can be
found.  The equilibration time within a metastable state is identified
by the time that the correlation function takes to stabilize to the
plateau value $C_{p}$. In the beta regime metastable states are
sampled in a quasi-ergodic fashion.  On this time scale, the
different four point correlation functions introduced in the previous
section can be then evaluated using constrained equilibrium measures
that select the relevant metastable states.  The set of constraints to
be introduced should insure that the relevant regions of configuration
space in the restricted measure coincide with the ones sampled by the
dynamics. The simplest possibility is to impose that in each region of
space the overlap with a well thermalized initial condition takes a
fixed value. We will suppose that this specification of the local
overlaps provides is a sufficient determination of the metastable
states and assume that configurations that have a fixed overlap close
to the plateau value with an equilibrium initial condition are sampled
(almost) ergodically. This hypothesis -sometimes called separability
\cite{separability}- can be checked directly in mean-field spin glass
systems and we believe to be valid in supercooled liquids.  In fact,
we expect it to apply every time that glassiness is caused by the
ruggedness of an energy landscape\footnote{On the contrary, we do not
  expect to apply in systems like kinetically constrained model, where
  the Hamiltonian is trivial. In this case the overlap does not give a
  sufficient determination of the metastable states \cite{FMP}.}.  In
the passage from dynamics to this quasi-equilibrium description we
loose of course the possibility of studying the time dependence of the
various quantities, that, as mentioned in the previous section, will be
expressed instead as 
 functions
 of the overlap in a {\it time
  reparametrization invariant} representation. In figure \ref{champ}
we illustrate how the four point dynamical susceptibility of the
spherical $p$-spin model \cite{FPchi4} looks like if we operate this
change of perspective.

Let us remark at this point that the use of a time independent
description of dynamical quantities has been widely used in the theory
of aging \cite{CuKu}, where reparameterization invariance emerges as
an asymptotic continuous symmetry at large times. It was then proposed
that this asymptotic zero mode could be used to characterize
fluctuations \cite{leticia} in glassy states. Being based on symmetry
considerations this theory is very general, and concerns 
features of both the beta and the alpha regimes.  Our theory, being
based on quasi-equilibrium considerations is less general and more
specific: it will enable to give a detailed description of the beta
regime, but it is limited to that. We will see however,
 when discussing
simulations, that looking at the data in reparametrization invariant
form is useful and inspiring also in the alpha regime.

In the rest of the paper we will concentrate on values of the
temperature close to $T_d$, and ignore the possibility of an ideal
glass transition at a lower temperature $T_K$.

We will concentrate on systems where either quenched disorder is
absent, like in real liquids, or, if disorder is present its effect is
weak and physical quantities can be evaluated to the leading order by
the ``annealed approximation''. This is a stronger property than the
usual self-averaging property of the free-energy and states that the
{\it partition function} has small sample to sample fluctuations.
Systems of this kind are often used to model structural glasses, and
include, among others, fully connected $p$-spin models, spin models on
diluted random graphs and finite range mean-field models in the Kac
limit.

\subsection{Effective Potential: a Landau field theoretical functional for the glass transition} 

According to the discussion of the previous section we can select
metastable states just choosing random equilibrium configurations
$S^0$ and restricting the Boltzmann measure to configurations that do
not differ too much from $S^0$. 
We achieve this fixing the local
overlaps $q_x(S,S^0)$ to preassigned values $p_x$ and defining
a constrained measure \cite{FPpot}
\begin{eqnarray}
\mu(S|S^0)=\frac {1}{Z[S^0,p_x]}e^{-\beta H(S)}\prod_x \delta(p_x-q_x(S,S^0)).
\label{measure}
\end{eqnarray}
For systems that are separable in the sense specified in the previous
section, the equilibrium metastable states are selected choosing in
all points of space $p_x=C_p$, profiles that deviate from this shape
allow to probe fluctuations.

The partition function ${Z[S^0,p_x]}$ is in fact directly related to
the probability of the overlap profile
\begin{eqnarray}
P(p_x|S^0)=e^{-\beta W(p_x,S^0)}=\frac{Z[S^0,p_x]}{Z}
\end{eqnarray}
where $Z$ is the unconstrained partition function. The large deviation
functional $W(p_x,S^0)$ depends on the choice of the overlap profile
but also on the choice of the reference configuration $S^0$ and on
quenched noise
in the case of
 disordered systems.  In our formalism any
dependence on $S^0$ quantifies the notion of ``self-generated
disorder'' often advocated in the physics of structural glasses
\cite{self-gen-dis}.  Previous studies have concentrated on the
average value of $W$ \cite{FPpot}.  Depending on the nature of system
under study, one can expect that the fluctuations of $W$ with respect
to $S^0$ and $J$ are more or less strong. For example, in a fully
connected model $W$ is a function of a single global overlap parameter
and self-averaging in the thermodynamic limit. Fluctuations decrease
as powers of the system size. We will see however that fluctuations of
the
correlation functions and fluctuations in the potential
can be related to each other.  The entire probability distribution of
$W$ is therefore relevant to a complete description of glassy systems.

In fact, the present formalism allows in principle to compute the
dynamic correlation functions that we have defined in the previous
section in {\it reparametrization invariant form}. To this scope we
introduce the generating function of the overlap $\Gamma(h_x|S_0)$
defined by
\begin{eqnarray}
e^{-\beta \Gamma(h_x,S^0)}=\int {\cal D}p_x\; e^{-\beta W(p_x,S^0)+\int dx h_x p_x} 
\end{eqnarray}
and define the static analogue of the correlations (\ref{4point}) in
presence of the field $h_x$ as
\begin{eqnarray}
&&G_{th}(x-y,h)= \mathbb{E} (\llbracket \langle p_x p_y\rangle- \langle p_x\rangle \langle p_y\rangle  \rrbracket )=
 \mathbb{E} (\llbracket \langle p_x p_y\rangle_c\rrbracket )
\nonumber \\
&&G_{het}(x-y,h)=  \mathbb{E} (\llbracket  \langle p_x\rangle \langle p_y\rangle  \rrbracket -\llbracket  \langle p_x\rangle  \rrbracket  \llbracket  \langle p_y\rangle  \rrbracket )=  \mathbb{E} (\llbracket  \langle p_x\rangle \langle p_y\rangle  \rrbracket_c)
\nonumber \\
&&G_{dis}(x,h)= \mathbb{E} (\llbracket  \langle p_x\rangle  \rrbracket  \llbracket  \langle p_y\rangle  \rrbracket )-
\mathbb{E} (\llbracket  \langle p_x\rangle  \rrbracket  )\mathbb{E} (\llbracket  \langle p_y\rangle  \rrbracket )
= \mathbb{E} (\llbracket  \langle p_x\rangle  \rrbracket  \llbracket  \langle p_y\rangle  \rrbracket )_c
\label{4pstat}
\end{eqnarray}
where we have denoted here by $\langle \cdot\rangle$ the equilibrium average in presence of $h_x$ and with a subscript ``$c$'' subtraction of the disconnected part. 
It is easy to check that the various correlations are related to the derivatives of the moments of the $\Gamma$ functional according to 
\begin{eqnarray}
&&G_{th}(x-y,h)= \mathbb{E} \left( \left\llbracket  \frac{\delta^2 \Gamma(h|S^0)}{\delta h_x\delta h_y} \right\rrbracket \right)
\nonumber \\
&&G_{het}(x,h)= \mathbb{E} \left(  \left\llbracket \frac{\delta \Gamma(h|S^0)}{\delta h_x}\frac{\delta \Gamma(h|S^0)}{\delta h_y}  
\right\rrbracket_c   \right)
\nonumber \\
&&G_{dis}(x,h)= 
\mathbb{E} \left(  \left\llbracket 
\frac{\delta \Gamma(h|S^0)}{\delta h_x}
\right\rrbracket 
\left\llbracket 
\frac{\delta \Gamma(h|S^0)}{\delta h_y}  
\right\rrbracket   \right)_c
\end{eqnarray}
If we fix the field $h_x$ in such a way that $\mathbb{E}(\llbracket \langle p_x\rangle\rrbracket )=q$ in all points of space we get the correlation functions as a function of $q$. 
We need then a method to compute the the cumulants of the functional $W$ or equivalently the ones of $\Gamma$.

\subsection{Effective potential and replicas}

It is interesting to compute both the average of the potential $W(q_x,S^0)$ and its fluctuations. 
The replica method gives us a simple framework to undertake this task. 
  As discussed many times \cite{FPpot}, the average
$W^{(1)}(p_x)=\mathbb{E} \llbracket W(p_x,S^0)\rrbracket $ can be computed considering
\begin{eqnarray}
Z_m(p_x)=\mathbb{E}\left( \llbracket  Z[S^0,p_x]^m \rrbracket \right) =\mathbb{E}
\left( \frac 1 Z\sum_{\{S^a\}_{a=0}^m} e ^{-\beta \sum_{a=0}^{m}
  H(S^a)}\prod_{a=1}^m \prod_x \delta( p_x-q_x(S^a, S^0) )\right)
\label{rep}
\end{eqnarray}
valid for integer $m$.\footnote{Thanks to the hypothesis of self-averageness of the partition function $1/Z\approx 1/\mathbb{E}(Z)$
  the average over disorder in (\ref{rep}) does not require additional
  care.}  Notice that here the total number of replicas, which
includes the reference configuration $S_0$ and the $m$ copies of the
constrained system, is ${\bf n}=m+1$.  The free-energy functional is
obtained from an analytic continuation to $m=0$, {\it i.e.} the total
number of replicas ${\bf n}$ tends to 1.
\begin{eqnarray}
W^{(1)}(p_x)=\left. -T\frac{\partial Z_m(p_x)}{\partial m}\right|_{m=0}-F
\end{eqnarray}
where $F$ is the average unconstrained free-energy of the system. 
Similarly one can get the second cumulants
\begin{eqnarray}
&& W^{(2)}_{het}(p_x,p'_x) = \mathbb{E}\big( \llbracket W(p_x,S^0)
  W(p_x',S^0) \rrbracket_c \big) = T^2 \left.\frac{\partial^2
    \log \mathbb{E} \big( \llbracket  Z[S^0,p_x]^{n_1} Z[S^0,p_x']^{n_2}
    \rrbracket \big) }{\partial n_1 \partial n_2}\right|_{n_1,n_2=0}\;,\\
&& W^{(2)}_{dis}(p_x,p'_x) = \mathbb{E}\big( \llbracket W(p_x,S^0)
  \rrbracket \llbracket W(p_x',S^0)\rrbracket \big)_c = T^2
  \left.\frac{\partial^2 \log \mathbb{E} \big( \llbracket
    Z[S^0,p_x]^{n_1} \rrbracket \llbracket Z[S^{0},p_x']^{n_2}
    \rrbracket \big)}{\partial n_1 \partial n_2}\right|_{n_1,n_2=0}\;,
\end{eqnarray}
where in the second equation we have exchanged the logarithm and the
average over the disorder, thanks to the annealed approximation.
Higher order cumulants can be analogously obtained through more
involved analytic continuations.

In order to unify the notation and treat all cases in parallel it is
convenient at this point to introduce the (formal) replica action
$S[Q_x]$ for ${\bf n}$ replicas for fixed values of their mutual
overlap $Q_{ab}(x)$ ($a,b=1,...,{\bf n}$) from:
\begin{eqnarray}
e^{-S[Q_x]} = \frac 1 Z  \mathbb{E} \sum_{\{S^a\}_{a=1}^{\bf n}} e^{-\beta
  \sum_{a=1}^{\bf n} H(S^a)}\prod_{a,b=e}^{\bf n} \prod_x
\delta(Q_{a,b}(x)-q_x( S^a,S^b))\;,
\end{eqnarray}
from which, integrating over some of the elements of the replica
matrix and fixing some others one can get the moments of $W$.  For
example one has that $W^{(1)}(p_x)$ can be computed by a replica
action with ${\bf n}=m+1$ replicas for $m\to 0$. 
Renumbering the replicas in a way that $a=0,1,...,m$ one has
\begin{eqnarray}
e^{-\beta m W^{(1)}(p_x)} = \int {\cal D}Q_{ab}(x)
e^{-S[Q_x]}\prod_{a=1}^{\bf m} \delta(Q_{0,a}(x)-p(x))\;,
\end{eqnarray}
Analogously the correlation functions can be computed from a replica action
with respectively ${\bf n}=n_1+n_2+1$ and ${\bf n}=n_1+n_2+2$ replicas
for $n_1,n_2\to 0$.
\begin{eqnarray}
&&W^{(2)}_{het}  (p_x,p'_x)  =-T\left.\frac{\partial }{\partial n_1\partial n_2}\right|_{n_1,n_2=0}
\log \int {\cal D}Q_{ab}(x) e^{-S[Q_x]}\prod_{a=1}^{n_1} \delta(Q_{0,a}(x)-p(x))\prod_{a=n_1+1}^{n_1+n_2} \delta(Q_{0,a}(x)-p'(x))\\
&& W^{(2)}_{dis}  (p_x,p'_x)  =-T\left.\frac{\partial }{\partial n_1 \partial n_2}\right|_{n_1,n_2=0}
\log \int {\cal D}Q_{ab}(x) e^{-S[Q_x]}\prod_{a=1}^{n_1} \delta(Q_{0,a}(x)-p(x))\prod_{a=n_1+1}^{n_1+n_2} \delta(Q_{0',a}(x)-p'(x)).
\end{eqnarray}
where in the first case we have renumbered the replicas in a way that
$a=0,1,...,n_1+n_2$ and in the second $a=0,0',1,...,n_1+n_2$.

We would like at this point to remind that in disordered mean-field
models there is a close relation between the Mode Coupling dynamical
transition and the shape of $W^{(1)}(q)$, which in that case is a
function of a single variable.  In fact, the transition temperature
$T_d$ looks as a spinodal temperature for the potential
$W^{(1)}$. This has a single minimum at low values of $q$ at high
temperatures, and develops a second minimum right at $T_d$ for the
value of the overlap $q=C_p$.

We argue that in a separable system, where the measure (\ref{measure})
correctly samples metastable states, this is the generic situation. If
metastability is found in some dynamical approximation, an
approximation for statics with the same physical content should lead
to the appearance of a secondary minimum in the average effective
potential corresponding to the constant profile $p_x=C_p$. Recent
analysis of MCT as a Landau expansion \cite{andreanov} on one hand and
reproduction of MCT results from replica Orstein-Zernike equations
\cite{szamel} on the other corroborate this point of view.

\section{The replica action close to $T_d$}
\label{replica-action}


We enter now in the core of our analysis, and we study fluctuations
for theories $S(Q)$ that at the level of homogeneous ({\it i.e.} space
independent) saddle point exhibit a dynamical phase transition at a
temperature $T_d$. This is associated to the appearance of a
horizontal inflection point at $C_p$ in the effective potential
$W^{(1)}(q)$, which becomes a minimum below $T_d$. As explained in
detail in \cite{FPpot} (see also \cite{remi}) this inflection point is
described by a ${\bf n}=1$ replica symmetric saddle point where
$Q_{ab}(x)=Q^d_{ab}=C_p$ for all $x$ and $a\ne b$. We wish to describe
overlap fluctuations for $T$ in the vicinity of $T_d$ and $p_x$ in the
vicinity of $C_p$.  A natural point of expansion of the action $S[Q]$
is the homogeneous saddle point just described.
We can then expand in $\epsilon=T-T_d$ and the difference of $Q_{ab}(x)$ with
$C_p$, $\phi_{ab}(x) =Q_{ab}(x)-C_p$ for $a\ne b$ and
$\phi_{aa}=0$. To the leading order one has a cubic theory
\begin{eqnarray}
S[Q,T]= &&S_0[Q^d,T_d]+\int dx\; \sum_{ab} \frac {\partial S[Q^d,T_d]}{\partial Q_{ab}(x) }\phi_{ab}(x)
+
 \sum_{ab} \frac {\partial^2 S[Q^d,T_d]}{\partial T\partial Q_{ab}(x) }\epsilon  \phi_{ab}(x)
\nonumber\\
&&+\int dx \; dy \;\sum_{ab;cd} \phi_{ab}(x)M_{ab;cd}(x,y)  \phi_{cd}(y)
\nonumber\\
&&+\int dx \; dy \;dz\; \sum_{ab;cd;ef} \Omega_{ab;cd;ef}(x,y,z)  \phi_{ab}(x)\phi_{cd}(y)\phi_{ef}(z). 
\end{eqnarray} 
The second term vanishes for ${\bf n}=1$ where $Q^d$ is a solution to
the saddle point equations. For generic ${\bf n}$ however this is not
the case, this will be a term of order ${\bf n}-1$ that has to be kept
in the expansion. To the lowest order in a gradient expansion and
rescaling the variables to reabsorb superfluous constants, the action
reads
\begin{eqnarray}
S[Q]=&& S_0[Q^d]+\int dx\; \sum_{ab}( A_{a,b}+ \epsilon ) \phi_{ab}(x)
\nonumber
\\
&&+\frac{1}{2} (\sum_{ab} \grad \phi_{ab}(x))^2 
+ \frac 1 2\sum_{ab;cd} \phi_{ab}(x)M_{ab;cd}  \phi_{cd}(x)
\nonumber
\\
&&+\sum_{ab;cd;ef} \Omega_{ab;cd;ef}  \phi_{ab}(x)\phi_{cd}(x)\phi_{ef}(x). 
\label{expa}
\end{eqnarray} 
Notice that the temperature couples linearly with $\phi_{ab}(x)$. This
is due to the choice of the point of expansion as the inflection point
at $T_d$.  The components of $A_{ab}$, the mass operator $M_{ab;cd}$
and of the bare vertex $\Gamma_{ab;cd;ef}$ reflect the symmetry of the
saddle point and should then depend only on the number indexes that
are equal or different. This immediately imply that the coefficients
$A_{ab}$ for $a\ne b$ should then be all equal $A_{ab}=A({\bf n})\sim
({\bf n}-1)A$, and that the quadratic form can be written as
\begin{eqnarray}
S_2[\phi]= \frac 1 2 \int dx\; \left( (\sum_{ab} \grad \phi_{ab}(x))^2+ 
 m_1 \sum_{ab}\phi_{ab}^2 +m_2 \sum_a(\sum_b \phi_{ab})^2 + m_3 (\sum_{ab} \phi_{ab})^2\right).
\label{quadra}
\end{eqnarray} 
The inclusion of all possible  replica symmetric cubic vertexes gives rise
to a cubic part \cite{DKT}: 
\beqa
& & S_3(\phi_{ab})=\int dx\; {\mathcal L}^{(3)} \label{cub}\\ 
{\mathcal L}^{(3)} & = & \frac{1}{6}\left[ \omega_1 \sum_{abc}
  \phi_{ab} \phi_{bc} \phi_{ca} + \omega_2 \sum_{ab}\phi_{ab}^3+
  \right. \nonumber \\
& + & \omega_3 \sum_{abc}\phi_{ab}^2\phi_{ac}+\omega_4 \sum_{abcd}
  \phi_{ab}^2 \phi_{cd}+\omega_5 \sum_{abcd} \phi_{ab}
  \phi_{ac}\phi_{bd}+ \nonumber \\
& + & \left. \omega_6 \sum_{abcd}\phi_{ab}\phi_{ac}\phi_{ad}+\omega_7
  \sum_{abcde}\phi_{ac}\phi_{bc}\phi_{de}+\omega_8
  \sum_{abcdef}\phi_{ab}\phi_{cd}\phi_{ef} \right]
\label{L3}
\eeqa
however, we will show that only the first two terms are relevant for
${\bf n}\to 1$.

Notice that the average potential within mean-field is evaluated by a
saddle point $\phi_{ab}(x)=\phi(x)$, which, inserted in (\ref{expa})
gives, to the lowest order in ${\bf n}-1$, 
\begin{eqnarray}
&&W^{(1)}(\phi_x)=W(0)+\int dx \; \frac{1}{2} (\nabla \phi)^2
  +\epsilon \phi +\frac{1}{2} m_1\phi^2 + g \phi^3\nonumber\\
&& g=\frac 1 6 (\omega_2-\omega_1)
\label{p1}
\end{eqnarray}
Since by hypothesis we have developed around the horizontal inflection
point for $\epsilon =0$, we must have $m_1=0$.
 
Of course, different choices for the point of expansion are
possible. In the region $T<T_d$ it is convenient to expand around the
replica symmetric saddle point $Q_{ab}=q_{EA}(T)$ at temperature $T$,
{\it i.e.} around the point that describe the minimum of the average
effective potential (\ref{p1}). This choice leads to an action like
(\ref{expa}) where all terms linear in $\phi$ are absent. The factor
$m_1$ is in this case non zero and proportional to
$\sqrt{-\epsilon}$. With this choice of the expansion point the
average potential reads $W^{(1)}(\phi_x)=W(0)+\int dx \; \frac{1}{2}
(\nabla \phi)^2 +\frac 1 2 m_1\phi^2 + g \phi^3$. In the following we
will use both expansions, without introducing separate notations for
the two.

Notice that we have written the expansion (\ref{expa}) for generic $D$
dimensional extended systems. However, as we will see in section
(\ref{f-s-s}), the same expansion can be used to describe finite size
corrections in fully connected disordered models and models on diluted
random graphs. In that case the various overlaps are global
quantities, the gradient term is absent and space integration is just
substituted by an overall multiplication by the system volume $N$.

\subsection{Quadratic free-energy fluctuations}
\label{qff}

In this section we discuss the correlation functions at the quadratic
(one loop) level, neglecting the cubic part of the action. We study
the fluctuations of the effective potential with respect to the choice
of the initial configuration and the choice of quenched disorder.  Our
task is to compute $W^{(2)}_{het}$ and $W^{(2)}_{dis}$. To this scope
we observe that the one loop order can be estimated as the saddle
point of the $n_1,n_2$ derivative of $S[\phi]$ where some of the
matrix elements are fixed. Let us start from
$W^{(2)}_{het}(\phi,\phi')$. In this case we need to consider a saddle
point of the action with ${\bf n}=1+n_1+n_2$ replicas where the
elements $\phi_{0,a}$ are fixed to $\phi_{0,a}(x)=\phi(x)$ for
$a=1,...,n_1$, $\phi_{0,a}(x)=\phi'(x)$ for $a=n_1+1,...,n_1+n_2$. In
presence of such constraints it is natural to look to a saddle point
which is symmetric with respect to all the permutations that leave
invariant the values of constraints, {\it i.e.} the group
$S_{n_1}\times S_{n_2}$ of independent permutations of the first group
and the second group of replicas among themselves. This is
parametrized in terms of three fields $\psi(x)$, $\psi'(x)$ and
$\psi_0(x)$ such that all the couples of replicas $a,b=1,...,n_1$ have
the same overlap $\phi_{ab}(x)=\psi(x)$, all the couples of replicas
$a,b=n_1+1,...,n_1+n_2$ have the same overlap $\phi_{ab}(x)=\psi'(x)$,
and all the couples of replicas $a=1,...,n_1$ $b=n_1+1,...,n_2$ have
an overlap $\phi_{ab}(x)=\psi_0(x)$. Inserting this ansatz, one
realizes that in the leading order in $n_1, n_2\to 0$ the equations
for $\psi$ (resp. $\psi'$) are independent from $\phi'$ and $\psi_0$
(resp. $\phi$ and $\psi_0$) and coincide with the ones that appear in
the computation of $W^{(1)}$.  In the limit of small $\epsilon$ the
solution is simply $\psi=\phi$, $\psi'=\phi'$ and $\psi_0=0$, which
gives
\begin{eqnarray}
W^{(2)}_{het}[\phi_x,\phi'_x]=\llbracket W^2(0|S^0)\rrbracket + A \int
dx\; (\phi(x)+\phi'(x))- (m_2+m_3) \int dx\; \phi(x)\phi'(x). 
\label{28}
\end{eqnarray}
This formula has a clear interpretation: the effect of the
heterogeneity in the reference configuration $S^0$ can be parametrized
in terms of a space dependent random free-energy shift $\alpha(x)$ and
a random temperature $\delta \epsilon(x)$ which couples linearly to
$\phi$.  This suggests that the potential $W(\phi|S^0)$ can be written
as
 \begin{eqnarray}
W(\phi|S^0) = \int dx\;\left[ \frac 1 2 (\nabla \phi)^2
  +(\epsilon+\delta \epsilon(x))\phi(x)+g\phi(x)^3 +\alpha(x)\right]. 
\label{rt}
\end{eqnarray}
The free-energy shift and the random temperature are Gaussian mutually
correlated fields with
\begin{eqnarray}
\llbracket \alpha(x)\,\alpha(y)\rrbracket & = & \llbracket
  W^2(0|S^0)\rrbracket \delta(x-y)/V\\
\llbracket \delta\epsilon(x)\,\delta\epsilon(y)\rrbracket &=&
-(m_2+m_3)\delta(x-y) \label{temp}\\
\llbracket \alpha(x)\,\delta\epsilon(y)\rrbracket &=& A\,\delta(x-y) 
\end{eqnarray}
where the consistency of the theory requires $m_2+m_3\leq 0$.  Formula
(\ref{rt}) is the central result of our paper, derived here at the
level of quadratic free-energy fluctuation.  The effective field
theory for the dynamic glass transition coincides with the one
describing the spinodal point of the Random Field Ising model and
therefore both problems are in the same universality class.  It is
well known that the leading singularities of random field theories in
perturbation theory are given by the tree diagrams \cite{PAS1}, or by
the formal solution of the stochastic differential equation
 \begin{eqnarray}
 - \Delta \phi +(\epsilon+\delta \epsilon(x))+3g\phi(x)^2=0. 
 \label{sde}
 \end{eqnarray}
In fact, even when this equation does not admit real solution, the
complex solutions gives rise to a perturbation series for physical
quantities with real coefficients.  Though the analysis of the
quadratic fluctuations give a strong hint about the validity of
(\ref{rt}), one could doubt that the inclusion of the vertexes in the
theory modifies this result.  As the matter of fact, in the next
section we will analyze in depth the perturbation theory for $T<T_d$
and confirm (\ref{rt}) to all orders in perturbation theory.

Let us now briefly turn our attention to $W^{(2)}_{dis}$.  In order to
compute this quantity  one may follow a route similar to the
computation of $W^{(2)}_{het}$. In this case however one can note that
in annealed models the replicas $0$ and $0'$ have zero overlap, and an
expansion around $q=C_p$ is not justified. Rather, one has that the
mutual overlap between the replicas labeled $1,...,n_1$ and the ones
label-led $n_1+1,...,n_1+n_2$ should be put to zero.  This leads to
decoupling between the two groups of replicas and $W^{(2)}_{dis}=0$ to
the leading order.  This is a remarkable result, showing that
fluctuations due to disorder are much less important than fluctuations
due to ``self-generated disorder'', seen here as heterogeneities in
the reference configuration.  We stress that the vanishing of
$W^{(2)}_{dis}$ at the quadratic level is consequence of the annealing
hypothesis and certainly would not be true in systems where disorder
fluctuations in the partition function have to be taken into
account. In our view this absence of dependence on quenched disorder
strongly supports the validity of long-range $p$-spin and similar
models as good mean-field models for the structural glass transition.

Let us now exploit (\ref{28}) to compute the overlap correlation
functions (\ref{4pstat}).  First of all we notice that at the tree
level calculation the potential $W$ and the generator $\Gamma$ are
related by
\begin{eqnarray} 
\Gamma_1[h]=\max_{q_x}W^{(1)}(q)-\int dx\; h_x q_x.
\label{w1}
\end{eqnarray}
This imply that the correlation function $G_{th}$ at the one loop level is given by
\begin{eqnarray} 
G_{th}(x-y)=\frac{\delta \Gamma_1[h]}{\delta h_x\delta h_y} =
\left(\frac{\delta W^{(1)}[q]}{\delta q_x\delta q_y}  \right)^{-1}.
\end{eqnarray}
We also notice that to the same accuracy
\begin{eqnarray} 
\Gamma_{het}^{(2)}[h,h']= W_{het}^{(2)}(q,q')
\end{eqnarray}
where $q$ and $q'$ are the maximizers of eq. (\ref{w1}) for field $h$
and $h'$ respectively.  A simple computation of the derivative of
$\Gamma_{het}^{(2)}$ shows that
\begin{eqnarray} 
G_{het}(x-y)=\frac{\delta \Gamma^{(2)}_{het}[h,h']}{\delta h_x \delta
  h_y'} =\int dx'\; dy'\;\frac{\delta W_{het}^{(2)} [q,q']}{\delta
  q_{x'} \delta q_{y'}'} G_{th}(x-x')G_{th}(y-y')
\label{double}
\end{eqnarray}
The same computation for $G_{dis}(x-y)$ would yield the same formula
with $W_{dis}^{(2)}$ at the place of $W_{het}^{(2)}$, but as we have
remarked $W_{dis}^{(2)}=0$ to the leading order.
This implies that no singularity of 
$G_{dis}$
can be detected in the quadratic theory. Beyond the gaussian
approximation,
the singularity of $G_{dis}$, if any, should be weaker then
the one of $G_{th}$ and $G_{het}$ and the possible 
presence of quenched disorder does not affect the universality class
of the system. 
Eq. (\ref{double})
shows that as soon as $\left(\frac{\delta W_{het}^{(2)} [q,q']}{\delta
    q_{x'}\delta q_{y'}'} \right)$ is non zero, the order of the
singularity in $G_{het}(x-y)$ is the double of the one of
$G_{th}(x-y)$. Notice that $\left(\frac{\delta W_{het}^{(2)}
    [q,q']}{\delta q_{x'}\delta q_{y'}'} \right)$ is precisely equal
to $\llbracket \delta\epsilon_x \delta\epsilon_y
\rrbracket=-(m_2+m_3)\delta(x-y)$.  We find the announced result that
the largest source of fluctuations in the system comes from the
heterogeneities in the initial condition. Its effect at the one-loop
level is to double the singularity due to thermal fluctuations. 

If we specify to the form (\ref{p1}) we find
\begin{eqnarray} 
\frac{\delta^2 W^{(1)} [q]}{\delta q_x\delta q_y}
=\delta(x-y)\left(-\Delta+6 g \phi(x) \right).
\end{eqnarray}
Fixing now a constant overlap profile in space $\phi$, we find that in
momentum space
\begin{eqnarray} 
&&G_{th}(k,\phi)=\frac{1}{k^2+6 g \phi}\\
&&G_{het}(k,\phi)=-\frac{(m_2+m_3)}{(k^2+6 g \phi)(k^2+6 g \phi)}. 
\label{gth1}
\end{eqnarray}
Both propagators are singular at $\phi=0$ and $k=0$, this corresponds
to the divergence of the fluctuations at $\epsilon=0$ and $q=C_p$. The
fluctuations for $\epsilon\ne 0$ can be obtains inserting for $\phi$ a
cut-off value of the order of the plateau $\phi_{EA}\sim
\sqrt{|\epsilon|/g}$ and
\begin{eqnarray} 
&& G_{th}(k)=\frac{1}{k^2+\sqrt{g|\epsilon|} }\\
&& G_{het}(k)=-\frac{(m_2+m_3)}{(k^2+\sqrt{g|\epsilon|} )(k^2+\sqrt{g|\epsilon|} )}. 
\label{gth2}
\end{eqnarray}

Notice that, within the present Gaussian approximation, the intensity
of critical temperature fluctuations $\llbracket \delta\epsilon_x
\delta\epsilon_y \rrbracket =-(m_2+m_3)\delta(x-y)$ can be measured
through the $\rho$ ratio
\begin{equation}
\rho \equiv \frac{G_{het}(k,\phi)}{G_{th}(k,\phi)^2}
\label{Rratio}
\end{equation}
at the critical point $\epsilon=0$, $\phi=0$. Beyond the Gaussian
approximation, while (\ref{gth1}) and (\ref{gth2}) provide a clear
indication that $G_{het}$ is more singular than $G_{th}$, it is not
clear to us if the relation $G_{het}\sim G_{th}^2$ continues to hold.  This
different scaling is an important result of our theory. Though our
derivation is restricted to the beta regime, we will see in section
(\ref{sec:simu}) that numerical simulations show that a different
scaling is also observed in the alpha regime.

\section{Perturbation Theory}
\label{pert}

We would like to confirm the description of fluctuations through the
potential (\ref{rt}) by perturbation theory. In order to have a well
defined point of expansion in perturbation theory, we assume $T<T_d$
and we expand around the replica symmetric minimum of the action
$Q_{ab}(x)=q_{EA}(T)$ for all $a,b$. Our starting point is then
\begin{eqnarray} 
S[\phi]= \int dx\;\frac 1 2\left( 
\sum_{ab} (\nabla \phi_{ab})^2 +
 m_1 \sum_{ab}\phi_{ab}^2 +m_2 \sum_a(\sum_b \phi_{ab})^2 + m_3
 (\sum_{ab} \phi_{ab})^2\right)+ {\mathcal L}^{(3)}[\phi_x] 
\label{expan}
\end{eqnarray} 
with ${\mathcal L}^{(3)}[\phi]$ given by (\ref{L3}) and
$m_1\sim\sqrt{-\epsilon}$.

\subsection{The bare propagators}

Let us now reobtain the results on the 4-point functions of section
(\ref{qff}) by studying the bare propagators of the replica field
theory, as first derived in \cite{CPR}. The analysis of a generic mass
matrix with replica symmetric structure has been performed long ago by
De Almeida and Thouless \cite{dat}. To analyze our case, we need to
transpose their results to the case ${\bf n}\to 1$.  There are three in
general distinct eigenvalues of the quadratic form named longitudinal
(L), replicon (R) and anomalous (A) in the current terminology.  For
future reference we remind that the longitudinal sector correspond to
fluctuations such that $\phi_{ab}^L=\phi$ independent of $a,b$ ($a\ne
b$), the replicon sector correspond to fluctuation such that $\sum_b
\phi_{ab}^R$ is vanishing for all $a$ and the anomalous sector is the
linear space of fluctuations orthogonal to the previous two.  In terms
of the parameters $m_1$, $m_2$ and $m_3$ of the quadratic form in
(\ref{expan}), the eigenvalues for generic ${\bf n}$ read
\beqa
\lambda_R(k) & = &k^2 + m_1 \\
\lambda_L(k) & = &k^2+ m_1+({\bf n}-1)(m_2+{\bf n} m_3) \\
\lambda_A(k) & = &k^2+ m_1+\frac{({\bf n}-2)}{2} m_2
\eeqa

The replicon $\lambda_R(0) = m_1\sim \sqrt{-\epsilon}$ is critical at
the transition.  Notice that for ${\bf n}=1$ the longitudinal
eigenvalue becomes
degenerate with the replicon and therefore it also become critical at
the transition.

It is already clear at this point that the replicon and longitudinal
sections will give the most singular contribution to the perturbation
theory. The propagator matrix of the theory in momentum space,
$G_{ab;cd}^{(0)}(k)=\left( k^2 +M \right)^{-1}_{ab;cd}$ has the same
replica symmetric structure as the mass matrix, so that we can write
\beq
G_{ab,cd}^{(0)}(k)=g_1\frac{(\delta_{ac}\delta_{bd}+\delta_{ad}\delta_{bc})}{2}+g_2
\frac{(\delta_{ac}+\delta_{ad}+\delta_{bc}+\delta_{bd})}{4}+g_3
\eeq
The coefficients $g_1,g_2$ and $g_3$ can be easily expressed in terms
of the eigenvalues of the mass matrix, or in terms of the $m$
parameters, the result for ${\bf n}=1$ is:
\beqa
g_1 & = & \frac{1}{ k^2+ m_1}
\\
g_2 & = &-\frac{ m_2} {m_2/2+ k^2+ m_1}\frac{1}{ k^2+ m_1}\sim \frac{1}{ k^2+ m_1} 
\\
g_3 & = &  \frac{1}{ k^2+ m_1}\left[
\frac{-(m_2+m_3)}{k^2+m_1}+\frac{ m_2} {m_2/2+ k^2+ m_1}
\right] 
\eeqa
The replica formalism naturally embeds the distinction among different
sources of fluctuations discussed in section (\ref{sec:corr}) and
allows to easily compute the propagators $G_{th}$ and $G_{het}$ of
section \ref{sec:corr} to the one loop order.  This can be done noticing that
$\mathbb{E} (\llbracket \langle \phi^2 \rangle \rrbracket )=
G_{ab,ab}$, $\mathbb{E}( \llbracket \langle \phi \rangle^2 \rrbracket)
= G_{ab,ac}$ where all the replica indexes are different one from
another
\beqa
&&G_{th}(k)=G_{ab;ab}-G_{ab;ac}=g_1-g_2\sim \frac{1}{k^2+ m_1} \\
&&G_{het}(k)=G_{ab;ac}=g_2+g_3\sim \frac{-( m_2+m_3) }
{(k^2 +  m_1)^2}.
\eeqa
Coherently with the results of section \ref{qff} the singularity of
$G_{th}$ is a single pole while the one of $G_{het}$ is a double
pole. Within the replica formalism the origin of the double
singularity stems from the degeneracy of the replicon and the
longitudinal eigenvalues for ${\bf n}\to 1$, which both become
critical at the transition.

\subsection{A perturbative derivation of the  Stochastic Equation}

We would like now to analyze the complete theory and show that the
leading singularities in perturbation theory coincide with the one
given by the stochastic differential equation (\ref{sde}). The
analysis we perform is similar to the analysis of random field models
as originally put forward by Parisi and Sourlas \cite{PAS1}.  This is
based on dimensional evaluation of the various vertexes of the theory
for ${\bf n}\to 1$ exploiting a change of basis that generalizes the
one suggested by Cardy in \cite{Cardy} for the RFIM.  We note first of
all that the leading singularities come from the replicon and
longitudinal modes, that become critical at the transition.  We
therefore concentrate on these modes, for which the matrix $\phi_{ab}$
is such that $\sum_b \phi_{ab}$ is independent of $a$. Symmetric
matrices with this property form a linear space of dimension $({\bf
  n}-1)({\bf n}-2)/2$.  We now describe fluctuation in a different
basis separating the fluctuating replica matrix in replica symmetric
part, independent of the indexes, plus a replica symmetry breaking
fluctuations and write, for $a\ne b$ and all points in space,
\begin{eqnarray} 
&& \phi_{ab} = \phi -\frac 1 2 \omega+ U_{ab}\; \omega+\chi_{ab}\\
&& \sum_{b}\chi_{ab}=0  \;\;\; \forall \; a\\
&& \sum_{a,b}\chi_{ab}U_{ab}=0
\label{pippo}
\end{eqnarray} 
Where we have defined $U_{ab}$ as a constant block matrix which has
all elements equal to zero except the ones for which $b=a+1$ and $a$
is odd or $b=a-1$ and $a$ is even, {\it i.e.}
\begin{eqnarray} 
U_{ab}=\left\{
\begin{array}{cl}
1 & {\rm if}\quad b=a-(-1)^a\\
0 & {\rm otherwise} 
\end{array}
\right.
\end{eqnarray} 
In this new basis we will be able, on the one hand, to perform the
${\bf n}\to 1$ limit directly in the action and, on the other hand, to
evaluate the scaling dimension of the different terms in the action in
order to keep only the most singular ones.  Notice that the
``vectors'' $\phi_{ab}$ defined by (\ref{pippo}) span the longitudinal
and replicon sector and that the matrices $\chi_{ab}$ verifying the
above relations form a linear space of dimension $({\bf n}-1)({\bf
  n}-2)/2-2$.  We observe now that
\begin{eqnarray} 
&&\sum_b \phi_{ab}=\omega +({\bf n}-1)(\phi -\frac 1 2 \omega)\\
&&\sum_{a,b} \phi_{ab}^2={\bf n} 2 \omega \phi
+\sum_{a,b}\chi_{ab}^2+{\bf n}({\bf n}-1)(\phi-\frac \omega 2 )^2.
\end{eqnarray} 
Neglecting all the terms that vanish in the ${\bf n}\to 1$ limit, the
quadratic part of the action reads
\begin{eqnarray} 
S_2=\int dx\; \left( 2 (\nabla \phi)(\nabla \omega)+\sum_{ab}(\nabla
\chi_{ab})^2 +m_1(\phi\omega +\frac 1 2 \sum_{a,b}\chi_{ab}^2)+\frac 1
2 (m_2+m_3)\omega^2 \right).
\label{s2}
\end{eqnarray} 
Let us now study the $m_1$-mass dimensions $D_\phi$, $D_\omega$ and
$D_\chi$ of the fields $\phi$, $\omega$ and $\chi_{ab}$. As usual we
impose that all terms in $S_2$ have the same dimension. We consider
the case in which $m_2$ and $m_3$ remain finite at the transition,
while $m_1\to 0$.
In this case we can write
\begin{eqnarray} 
&&D_\omega =D_\phi+ 1\\
&&D_\chi=D_\phi+ \frac 1 2
\end{eqnarray} 
Let us now analyze the vertexes. Expressed in the new
variables the first four vertexes of (\ref{L3}) read:
\begin{eqnarray} 
\sum_{a,b,c}\phi_{ab}\phi_{bc}\phi_{ca}
 =&&
({\bf n}-1)({\bf n}-2)  (\phi-\frac \omega 2 )^3-
3\phi\sum_{ab}\chi^2_{ab}+
3{\bf n} ({\bf n}-2)\omega \phi^2+
{\rm Tr}\; \chi^3
\nonumber\\
&&+
\omega {\rm Tr}\chi^2 U-
\frac 3 2 \omega\sum_{ab}\chi_{ab}^2+
3 \phi \omega^2 {\bf n}({\bf n}-2)+
\omega^3 {\bf n}(\frac 3 4 {\bf n} -2)
\\
\sum_{ab} \phi_{ab}^3=&&
{\bf n}({\bf n}-1) (\phi-\frac \omega 2 )^3+
3\phi\sum_{ab}\chi^2_{ab}+
3{\bf n} \omega \phi^2+
\sum_{ab} \chi_{ab}^3
\nonumber \\&&+
3 \omega\sum_{ab} \chi_{ab}^2U_{ab}-
\frac 3 2 \omega\sum_{ab} \chi_{ab}^2+
6{\bf n }\omega^2 \phi
+{\bf n}\frac {\omega^3}{ 4} 
\\
\sum_{a,b,c}\phi_{ab}^2\phi_{ac}=&& 
(\omega+({\bf n}-1)(\phi-\frac 1 2 \omega))
\left( 
{\bf n} 2\omega \phi +\sum_{a,b}\chi_{ab}^2+{\bf n}({\bf n}-1)(\phi+\frac \omega 2 )^2 
\right)
\\
\sum_{a,b,c,d}\phi_{ab}^2\phi_{cd}= &&
n\sum_{a,b,c}\phi_{ab}^2\phi_{ac} 
\end{eqnarray} 
all the other combinations that appear in $S_3$, i.e. vertexes 
5 to 8, give just rise to
terms proportional to $\omega^3$ plus terms that vanish for ${\bf
  n}\to 1$. Notice that all the terms
proportional to $\phi^3$ vanishes for ${\bf n}\to 1$. In the new
basis, the mass
dimensions of the vertexes that survive for ${\bf n}\to 1$ are
\begin{eqnarray} 
\omega \phi^2 , \; \phi\sum_{ab}\chi_{ab}^2 &\to& 3 D_{\phi}+1
\label{list}
\\
 \sum_{ab}\chi_{ab}^3,\;  {\rm Tr}\; \chi^3 &\to&  3 D_{\phi}+\frac 3 2 \\ 
\omega \sum \chi_{ab}^2, \; \omega {\rm Tr }\chi^2 U,\; \omega^2 \phi &\to&  3 D_{\phi}+ 2 \\ 
\omega^3 &\to& 3 D_{\phi}+ 3 
\end{eqnarray} 
The leading singular behavior in perturbation theory of the theory for
$m_1\to 0$ is dictated by the the first two vertexes (\ref{list}),
which are the ones of lower dimension.  Neglecting therefore the
subleading vertexes we find that we can write the action as
\begin{eqnarray} 
S=\int dx\; \frac 1 2 (m_2+m_3) \omega^2 + \omega\left( 
-\Delta \phi + m_1\phi+3 g\phi^2
\right)+\frac 1 2\sum_{a,b}\chi_{ab} \left( 
-\Delta + m_1+6 g\phi
\right)\chi_{ab}
\label{fin}
\end{eqnarray} 
where $g=\omega_2-\omega_1$. We observe that the matrix field
$\chi_{ab}(x)$ has $({\bf n}-1)({\bf n}-2)/2-2\to -2$ independent
components and becomes equivalent to a couple of anticommuting fermion
fields $\chi(x)$ and ${\bar\chi(x)}$ for ${\bf n}\to 1$. Equivalently,
we can observe that the explicitly integration over the $\chi_{ab}$
fields gives rise to
\begin{eqnarray} 
\det \left(-\Delta  + m_1+6 g\phi(x) \right)^{1-({\bf n}-1)({\bf
    n}-2)/4}\mathop{\longrightarrow}_{{\bf n}\to 1} \det \left(-\Delta
+ m_1+6 g\phi(x) \right). \label{det}
\end{eqnarray} 
We finally recognize in the action (\ref{fin}) a Parisi-Sourlas 
supersymmetric theory associated with the stochastic equation 
\begin{eqnarray} 
-\Delta \phi(x) + m_1\phi(x)+3 g\phi(x)^2+\delta\epsilon(x)=0
\label{se}
\end{eqnarray} 
where $\delta\epsilon(x)$ is a gaussian field with variance 
\begin{eqnarray} 
\llbracket \delta\epsilon(x)\delta\epsilon(y)\rrbracket =-(m_2+m_3)\delta (x-y).
\label{e2}
\end{eqnarray}
If we impose that all the terms in the action have the same scaling
dimension we find $D_\phi=1$ and in $D$ spatial dimension
 the action has dimension $-\frac D
2+4$. It is well known that Parisi-Sourlas actions present a
supersymmetry that leads to the phenomenon of dimensional
reduction. The perturbation theory of the system in a random field in
dimension $D$ coincides with the one of a pure system in dimension
$D-2$. Coherently the upper critical dimension of the theory is
promoted to 8 from the value 6 of the pure $\phi^3$ theory. 
 
From the (formal) solution of eq. (\ref{se}) $\phi(x)$, we can obtain
the correlation functions $G_{th}$ and $G_{het}$ through linear
response theory, using (\ref{e2})
\begin{eqnarray}
&&G_{th}(x-y)=\llbracket
  \frac{\delta\phi(x)}{\delta\epsilon(y)}\rrbracket =
  \frac{-1}{m_2+m_3} \llbracket
  \phi(x)\times\delta\epsilon(y)\rrbracket 
\nonumber\\
&&G_{het}(x-y)= \llbracket \phi(x)\phi(y)\rrbracket_c. 
\end{eqnarray} 

The derivation leading to (\ref{fin}), (\ref{det}), (\ref{se}),
(\ref{e2}) is valid within perturbation theory. Its expression should
be considered as a formal writing valid within the perturbative
context.  In fact, due to the cubic nature of the potential,
eq. (\ref{se}) admits a real solution only if $\delta\epsilon(x)$ is
sufficiently small in absolute value in all points of space.  For a
gaussian field $\delta\epsilon(x)$ this condition is violated with
probability one in an infinite space. The consequence of that is that
the perturbation series should be divergent.  In fact, the
perturbation theory is formally identical to the one of branched
polymers considered in \cite{PAS2}, with the crucial difference that
in polymer's case the coupling constant $g$ is purely imaginary
\cite{fisher}, while in the present case it is real.  In the branched
polymer case the perturbative series is resummable thanks to the fact
that its terms have alternating signs. Here, all terms have the same
sign, the resulting series is badly divergent and it seems hardly
resummable. Recent work has used ``Exact Renormalization Group''
\cite{erge} methods to compute the exponents of the thermodynamic
transition of the RFIM \cite{tarjus}. It is not clear to us if these
methods can be useful to study the spinodal point.

A fast way of realizing that perturbation theory should be divergent
comes from considering equation (\ref{se}) in the homogeneous limit of
space independent quantities. As we will describe in the next section,
this allows to describe finite size scaling in mean-field models. In
this case the equation reads
\begin{eqnarray} 
 m_1\phi+3 g\phi^2+\delta\epsilon=0
\label{se-mf}
\end{eqnarray} 
where now the variance of $\delta\epsilon$ is given by
\begin{eqnarray} 
\llbracket \delta\epsilon\delta\epsilon\rrbracket =-\frac{(m_2+m_3)}{N}.
\end{eqnarray} 
The correlation function $G_{het}$ can be formally evaluated from the
solution $\phi^*=\frac{-m_1+\sqrt{m_1^2-12g \delta \epsilon }}{6 g}$
as:
\begin{eqnarray} 
G_{het}(\epsilon, N)=\frac{1}{N} \left( \llbracket (\phi^*)^2
\rrbracket -  \llbracket \phi^* \rrbracket^2 \right)
\label{corr}
\end{eqnarray} 
This correlation can be evaluated in an expansion in $\delta\epsilon$,
giving rise to well defined series with real coefficients. However it
is clear that the series cannot be convergent as the averages in
(\ref{corr}) receive an imaginary contribution from the square root
in the region $\delta\epsilon>m_1^2/12g$.

Putting these problems of convergence aside, one can ask if, for
typical disorder realization $\delta\epsilon(x)$, the stochastic
equation (\ref{se}) could be used beyond perturbation theory in a
description of the barrier jumping processes in a reparametrization
invariant way. While we do not have a definite answer in general we
will discuss the consequence of this hypothesis for mean-field finite
size scaling in the next section.

\section{Finite size scaling around mean field}
\label{f-s-s}

\subsection{The beta region}

The theory that we have developed can be easily generalized to
describe finite size scaling in mean-field systems which have a
genuine MC dynamical phase transition in the thermodynamic
limit. These include disordered spin models like p-spin and Potts
defined on fully connected or finitely connected random graphs, and
problems that appear in computer science and information theory like
the K-SAT or XOR-SAT problem or error correcting codes.  For large but
finite sizes $N$ the dynamical transition is cut-off. Finite size
scaling should describe the cross-over to criticality for $N\to\infty$
and $\epsilon\to 0$.

A phenomenological theory of dynamic finite size scaling for
disordered mean field models has been first proposed in
\cite{SBBB}. In order to interpret numerical results, in
Ref.~\cite{SBBB} it was assumed a sample dependent critical
temperature and the Harris criterion was used to derive scaling
variables and exponents. Our results, using a fundamental theoretical
description, confirm and rationalize that analysis 
as far as
reparametrization invariant quantities are concerned. 
As discussed
above, the origin of the random temperature term is in our theory a
consequence of dynamical heterogeneity rather than of quenched
disorder.

To analyze mean-field finite size scaling we follow the lines drawn in
the previous sections, except that in this case overlaps do not
display space, that is we use observables integrated over the whole
system.  The relevant replica action is identical to the one discussed
in section \ref{replica-action} without the gradient term and with
space integration substituted by an overall volume factor $N$.

Repeating the analysis of the replica action that led to (\ref{rt}) and
(\ref{se}) in this case, we get a description in terms of a single
variable effective potential $W(\phi|S_0)$ describing the total
overlap fluctuations around the plateau value, $\phi=q-C_p$
\begin{eqnarray}
W(\phi|S^0)=W(0|S^0)+N\left(
[\epsilon+\delta \epsilon] \phi +g \phi^3 +\alpha
\right)
\label{77}
\end{eqnarray}
where in this case both $\delta\epsilon$ and $\alpha$ are gaussian
covaring variables of order $1/\sqrt{N}$. In terms of the parameter of
the replica field theory, $g=\frac 1 6 (\omega_2-\omega_1)$,
$\llbracket \alpha\alpha \rrbracket=\frac 1 N \llbracket W^2(0|S_0)
\rrbracket$, $\llbracket \alpha\delta\epsilon \rrbracket=\frac 1 N A$,
$\llbracket \delta\epsilon^2 \rrbracket=-\frac 1 N (m_2+m_3)$.
Correspondingly we get the Parisi-Sourlas action
\begin{eqnarray}
S=N\left(
\frac{m_2+m_3}{2} \omega^2+\omega(\epsilon+3g\phi^2)+\chi 6 g \phi {\bar\chi}
\right).
\label{PS78}
\end{eqnarray} 
From formula (\ref{PS78}) the properties of finite size scaling readily
follow from dimensional analysis.  We are interested to the behavior
of the various observables in the critical cross-over region for
$N\to\infty$ and $\epsilon\to 0$. This is the region of variables such
that all terms in the action are of order one,
namely $\phi\sim\sqrt{|\epsilon|}$, $\omega\sim \epsilon\sim N^{-1/2}$
and $\chi,{\bar \chi}\sim N^{-3/8}$. This allow to 
identify as scaling variables $x=\phi N^{1/4}$ and $y=\epsilon
N^{1/2}$ in quantities that depend on size, temperature and overlap. 
Mutating the results of section \ref{qff} on the behavior
of quadratic fluctuations, we find that, for $\phi\to 0$ and
$N\to\infty$, $\chi_{th} \sim \frac{1}{\sqrt{|\epsilon|}}$ and
$\chi_{het}\sim \frac{1}{{|\epsilon|}}$.  Trading $N$ for $\epsilon$, 
this implies that in the
cross-over region
\begin{eqnarray}
&&\chi_{th}(\phi,\epsilon,N)= N^{1/4} f_{th}( \phi N^{1/4},\epsilon
N^{1/2}) \\ &&\chi_{het}(\phi,\epsilon,N)=N^{1/2} f_{het}( \phi
N^{1/4},\epsilon N^{1/2}).
\label{fss}
\end{eqnarray}
In order to match the singularities for $N\to\infty$, the scaling
functions $f_{th}(x,y)$ and $f_{het}(x,y)$ verify $f_{th}(0,y)\sim
\frac{1}{\sqrt{y}}$, $f_{th}(x,0)\sim \frac{1}{x}$ and
$f_{het}(0,y)\sim \frac{1}{y}$, $f_{het}(x,0)\sim \frac{1}{x^2}$ for 
$x\to +\infty$ or $y\to +\infty$, while
a finite limit should be expected at small values of the two
arguments.

It is interesting to study the behavior of the ratio
$\rho=\chi_{het}/\chi_{th}^2$ in the scaling window.  In the
$N\to\infty$ limit this ratio can be related to the variance of the
random field by the relation: $\lim_{\phi\to 0}\lim_{N\to\infty}
\rho(\epsilon=0,\phi,N)=-(m_2+m_3)$ where the limits should be taken
in the order.  In the scaling window on the other hand
$\rho(\epsilon,\phi,N)=f_{het}(x,y)/f_{th}(x,y)^2$. Consistency
requires that
$\lim_{y\to\infty}f_{het}(0,y)/f_{th}(0,y)^2=-(m_1+m_2)$.
As we have observed in the previous section the perturbative series of the
scaling functions are badly divergent and can hardly be
computed analytically. Thus, this is as far as we can go from the perturbative
analysis of (\ref{PS78}).  Numerical verification of the scaling forms
(\ref{fss})
 will be the object of next section.

\subsection{The alpha regime: some conjectures}
\label{a-conj}

In the deep alpha regime, where $\phi<0$ is of order one in absolute value,
quasi-equilibrium in the sense we were using so far does not hold, the
configuration space with average correlation $C<C_p$ is not sampled
ergodically and our theory does not apply. However, this does not make
it less interesting to look at the time reparameterization
invariant part of the fluctuations and on the contrary calls for new 
theoretical ideas to be developed. One can hope that finite
size scaling of mean field models is a simple enough setting to put
forward hypotheses to be tested in the general case. 
This section will be then by nature much more conjectural and
qualitative than the previous ones.  

In order to understand finite size scaling in the alpha region we can
conjecture continuity with the behavior in the scaling regime
$\phi\sim N^{-1/4}$ and quote the results of the analysis of dynamic
Gaussian fluctuation theory developed in \cite{BBBKMR}. This predicts
that the peak $\chi_4^*$ of the dynamic susceptibility 
as a function of time, lies deep in the alpha region and 
scales as $\frac{1}{\epsilon^2}$.\footnote{In
  \cite{BBBKMR} this behavior was found to hold only for conservative
  dynamics. Later analysis showed that the same behavior
  also holds generically even in absence of dynamically conserved quantities
  \cite{giulio}.}. 
If we assume that this scaling holds in the whole regime $C<C_p$, we
can match the form (\ref{fss}) if we suppose that the function
$f_{het}(x,y)$ behaves quadratically for large negative values of
$x$. In this way $\chi_{het}(\phi,\epsilon,N)=N^{1/2} f_{het}( \phi
N^{1/4},\epsilon N^{1/2})\simeq N\phi^2g_{het}(\epsilon N^{1/2})$ for
$-\phi N^{1/4}>>1$, where the function $g_{het}(y)$ should behaves as
$y^{-2}$ for large argument and take finite value for $y=0$.  This
implies that in the alpha region one one has that for $\epsilon=0$,
$\chi_{het}=O(N)$. If we also assume that $f_{th}(x,y)$ behaves
quadratically for large negative $x$, we get
$\chi_{th}(\phi,\epsilon,N)=N^{1/4} f_{th}( \phi N^{1/4},\epsilon
N^{1/2})\simeq N^{3/4}\phi^2g_{th}(\epsilon N^{1/2})$; which implies a
$N^{3/4}$ scaling of $\chi_{th}$ for $\epsilon=0$ and a behavior as
$\epsilon^{-3/4}$ for small positive $\epsilon$ and $N\to\infty$. A
linear behavior $f_{th}(x,y)=x h_{th}(y)$ would give $\chi_{th}\sim
N^{1/2}$ for $\epsilon=0$ and $\chi_{th}\sim \epsilon^{-1}$ for
small $\epsilon$ and $N\to\infty$.

Away from criticality, a simple dynamical scenario can be conjectured for the alpha
relaxation in the barrier  dominated  region $\epsilon<0$ and
$N\to\infty$. We make a crucial assumption that in a first
approximation in the region $\epsilon\leq 0$, the reparameterization
invariant fluctuations in the alpha region are equivalent to the ones
of a simple barrier jumping process, as described for example by the Langevin
equation
\begin{eqnarray}
\frac{dq}{ds}=-W'(q|S^0)+\eta
\label{mf-langevin}
\end{eqnarray}
in the weak noise limit $\eta \ll 1$. The waiting time before a jump
is a random variable whose typical values are much larger then the
time required to pass from $C_p$ to $0$ when the jump occurs.
This leads to a bistable behavior 
that naturally gives rise to $O(1)$
fluctuations of the overlap. At any given instant of time one has 
just a small probability of finding
$C(t)$ to be different from $C_p$ or $0$.  Neglecting this probability
we find a general form of the total susceptibility $\chi_4(q)$, which
is independent of the barrier. Suppose now to fix time in a way that
for each value of the system size $C_{av}=\llbracket \langle C\rangle
\rrbracket$.  The histogram of $C$ for different trajectories and
initial conditions is approximately given by
\begin{eqnarray}
P(C|C_{av})=\frac{C_{av}}{C_p} \delta(C-C_p)+(1-
\frac{C_{av}}{C_p} )\delta(C)
\label{pc}
\end{eqnarray}
The direct computation of $\chi_4(C_{av})$ using (\ref{pc}) leads to the
simple expression
\begin{eqnarray}
\chi_4(C_{av})/N=C_{av} (C_p-C_{av}).
\label{chi-alpha}
\end{eqnarray}
This form is independent of the barrier and valid in the whole
region $T< T_d$. As we will see in the next section, these
hypotheses provide a good (but still not perfect) description of
numerical results even at $\epsilon=0$.
Notice that in deriving (\ref{chi-alpha}) we have just used 
bistability and not the detailed Langevin equation (\ref{mf-langevin}). 

One can then try to use similar ideas to describe the alpha relaxation
in finite dimensional systems.  For example, numerical simulations in
\cite{KDS-pnas} suggest that bistability as described by (\ref{pc})
approximately holds for the probability of the local overlap $q_x(t)$
coarse grained on scales smaller than the correlation length. It would
be then tempting to write an equation analogous to
(\ref{mf-langevin}), with an additional Laplacian term to describe the
decay of the overlap. Unfortunately the phenomenology of such an
equation \cite{theoryofthecondensationpoint}, which describes the
competition between different phases in presence of disorder, is
rather distant from the one of supercooled liquids. In particular such an equation
would predict nucleation events as they could be expected in liquids, but
also the fast growth of supercritical nuclei of low overlap that
cannot be expected. 

\section{Simulations}
\label{sec:simu}

In order to test the theoretical predictions derived above, we have
simulated the 3-spin model on a random regular graph, {\it i.e.}\ a
random graph of fixed connectivity $z$. This model involves $N$ Ising
spins $S_i=\pm 1$ interacting through the Hamiltonian
\begin{eqnarray}
H[S]=-\sum_{\mu=1}^M J_\mu S_{i_1^\mu}S_{i_2^\mu}S_{i_3^\mu}
\end{eqnarray}
where the indexes $i_1^\mu, i_2^\mu, i_3^\mu$ are chosen at random in
such a way that each spin participates exactly to $z=3M/N$
interactions and the couplings $J_\mu$ are independent random
variables taking values $\pm 1$, with $\mathbb{P}(J_\mu=1)=r$. The
properties of the model in thermodynamic limit are well known
\cite{FedeAndreaPRB}. In particular if $z \ge 4$ the model exhibit a
MC like dynamical phase transition at temperature $T_d$ and a Kauzmann
ideal glass transition at a lower temperature $T_k$; both transition
temperatures depend on $z$, but not on $r$. In fact it is well known
that the thermodynamic and dynamic properties of the model are
independent of the choice of $r$ for temperatures $T>T_k$. In this
range of temperature the model is annealed: that is, denoting $Z_J$
the partition function for a given disorder realization $J$, one has
that $\lim_{N\to\infty} \frac 1 N\mathbb{E} \log Z_J=
\lim_{N\to\infty} \frac 1 N\log \mathbb{E}Z_J$.  This property is true
in particular for the \emph{symmetric model} with $r=1/2$ and for the
\emph{gauged model} defined by the Nishimori condition
$r=(1-e^{-2\beta})^{-1}$ \cite{Nishi}.  Although the two versions of
the model (the symmetric and the gauged one) have the same
thermodynamical behavior for $T>T_k$, they may show very different
finite size effects.  In the following we study both versions.

We choose to simulate the case $z=8$. The analytic solution to the
model \cite{FedeAndreaPRB} predicts, in the thermodynamic limit, a
dynamical temperature $T_d=1.3420(5)$ and the plateau value at $T_d$
equal to $C_p=0.750(5)$. All the results shown in the following have
been obtained at the dynamical critical temperature $T_d$.

\subsection{Simulations in the beta regime}

We start by showing results obtained with the symmetric model
($r=1/2$).  We have simulated at the dynamical critical temperature
$T_d$ systems of size ranging from $N=60$ to $N=150$, with a number of
samples $N_S$ such that $N \times N_S = 1.8 \cdot 10^6$. For each
sample, we have obtained 2 independent equilibrium configurations
(with the use of the parallel tempering algorithm), and, from each of
these, we have evolved 2 independent trajectories with different
thermal noises: this allow us to compute all the three different
fluctuations $\chi_{dis}$, $\chi_{th}$ and $\chi_{het}$
(see below the
detailed explanation for the gauged model).

\begin{figure}[!ht]
\begin{center}
\includegraphics[width=0.6\textwidth]{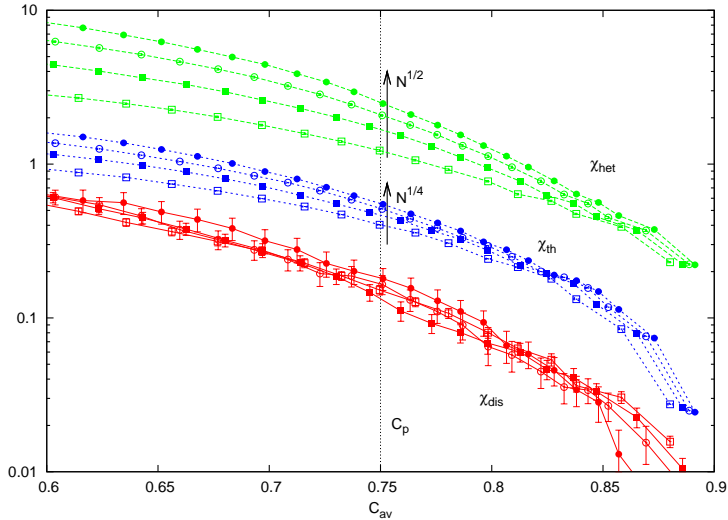}
\caption{Three different susceptibilities $\chi_{dis}$, $\chi_{th}$
  and $\chi_{het}$ (from bottom to top) {\it versus} average
  correlation $C_{av}$ for the 3-spin symmetric model. The $\chi_{th}$
  data have been divided by a factor 3 in order to avoid data
  overlap. The vertical line marks the analytical value of $C_p=0.75$.
  System sizes are $N=60,90,120,150$ (from bottom to top for each
  susceptibility).}
\label{rawDataSymm}
\end{center}
\end{figure}

In Fig.~\ref{rawDataSymm} we plot these three susceptibilities
$\chi_{dis}$, $\chi_{th}$ and $\chi_{het}$ measured in the symmetric
model ($r=1/2$) for sizes $N=60,90,120,150$.  In the symmetric model,
we can not study larger sizes due to the very large thermalization
times.  Nonetheless, even for these relatively small sizes, we can
clearly see the very different size dependence of the susceptibilities
at the critical point, $C_{av}=C_p$. As expected from the discussion
in section \ref{qff},  $\chi_{dis}$ is
practically size independent within error bars, while $\chi_{th}$ and
$\chi_{het}$ are well compatible with the predicted scaling laws,
$N^{1/4}$ and $N^{1/2}$ respectively.  Unfortunately data are plagued
by severe finite size effects even in the beta region, $C_{av} > C_p$.

In order to reach a clearer conclusion about the scaling laws in the
beta regime and at the critical point, we need to study larger systems
and this is the reason for using the gauged model, where an
equilibrium configuration can be generated without the long
thermalization process \cite{sem,flo}.  Indeed on the Nishimori line,
for a given interaction graph, one can first choose an arbitrary spin
configuration $S^0$, and then fix the coupling $J_\mu$ such that $S^0$
is an equilibrium configuration. This can be done by choosing the
couplings as independent random variables taking values $\pm 1$ with
the following probability:
\begin{eqnarray}
\mathbb{P}(J_\mu|S^0)=\frac{\exp(-\beta J_\mu
S_{i_1^\mu}^0S_{i_2^\mu}^0S_{i_3^\mu}^0 )}{2\cosh(\beta)}.
\end{eqnarray}
Since $S^0$ can be arbitrary, it is customary to set $S^0_i=1$ for all
$i$, convention that we will adopt in the following.  As a drawback of
the method we note that since we generate at the same time initial
configuration and quenched couplings, we cannot disentangle the
contributions of these two sources of noise in the fluctuations. With
this method we will therefore be able to compute $\chi_{th}$ and the
sum of $\chi_{het}+\chi_{dis}$ but not each term separately. Since we
expect the contribution due to the quenched disorder ($\chi_{dis}$) to
be small, we will improperly call $\chi_{het}$ the latter sum. Notice
that besides the choice of the random coupling $J_\mu$, an additional
source of quenched disorder comes from the choice of the random
graph. It is known however that local properties of random regular
graphs are self-averaging and consequently the effect of topology
fluctuations are even smaller.

We have simulated at the dynamical critical temperature $T_d$ systems
of size ranging from $N=150$ to $N=2400$, with a number of samples
$N_S$ such that $N \times N_S = 3.7 \cdot 10^7$. 
In order to be able to measure both $\chi_{th}$ and $\chi_{het}$, for
each sample $\alpha=1,...,N_S$ we have simulated two independent
trajectories $s=1,2$ starting from the same
equilibrium configuration and evolving with different thermal noises. We then for each sample and 
each trajectory measure the correlation functions
$C_{\alpha,s}(t)=\frac 1 N \sum_iS_i^{\alpha}(0)S_i^{\alpha,s}(t)$,
where the initial state is equal for the two trajectories. The
suceptibilities are then estimated as
\begin{eqnarray}
\chi_{th} &=&\frac{1}{2 N_S}\sum_{\alpha,s}(C^{\alpha,s})^2
-\frac{1}{ N_S}\sum_{\alpha,s}C^{\alpha,1}C^{\alpha,2}\\
\chi_{het} &=&
\frac{1}{ N_S}\sum_{\alpha,s}C^{\alpha,1}C^{\alpha,2}
-\left(\frac{1}{2 N_S}\sum_{\alpha,s}(C^{\alpha,s})\right)^2.
\end{eqnarray}

\begin{figure}[!ht]
\begin{center}
\includegraphics[width=0.6\textwidth]{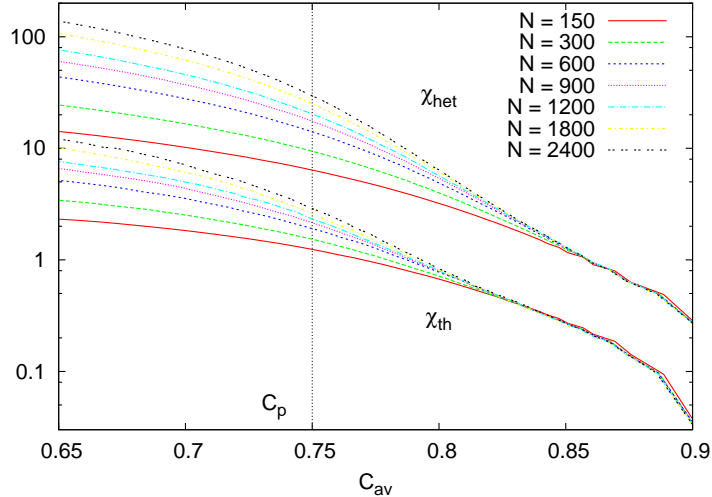}
\caption{Thermal susceptibility $\chi_{th}$ (lower data) and
  heterogeneity susceptibility $\chi_{het}$ (upper data) {\it versus}
  average correlation $C_{av}$ for the 3-spin gauged model. The
  vertical line marks the analytical value of $C_p$.}
\label{rawDataGauged}
\end{center}
\end{figure}

\begin{figure}[!ht]
\begin{center}
\includegraphics[width=0.49\textwidth]{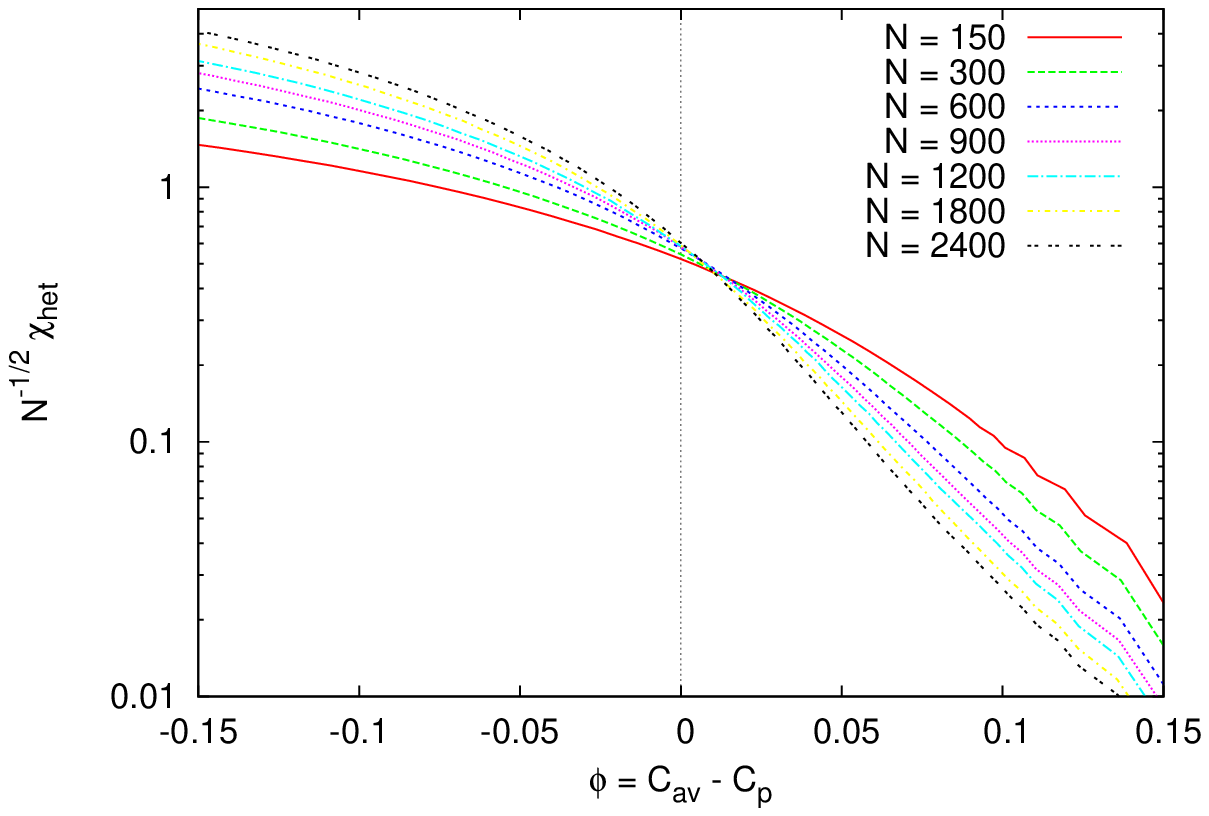} 
\includegraphics[width=0.49\textwidth]{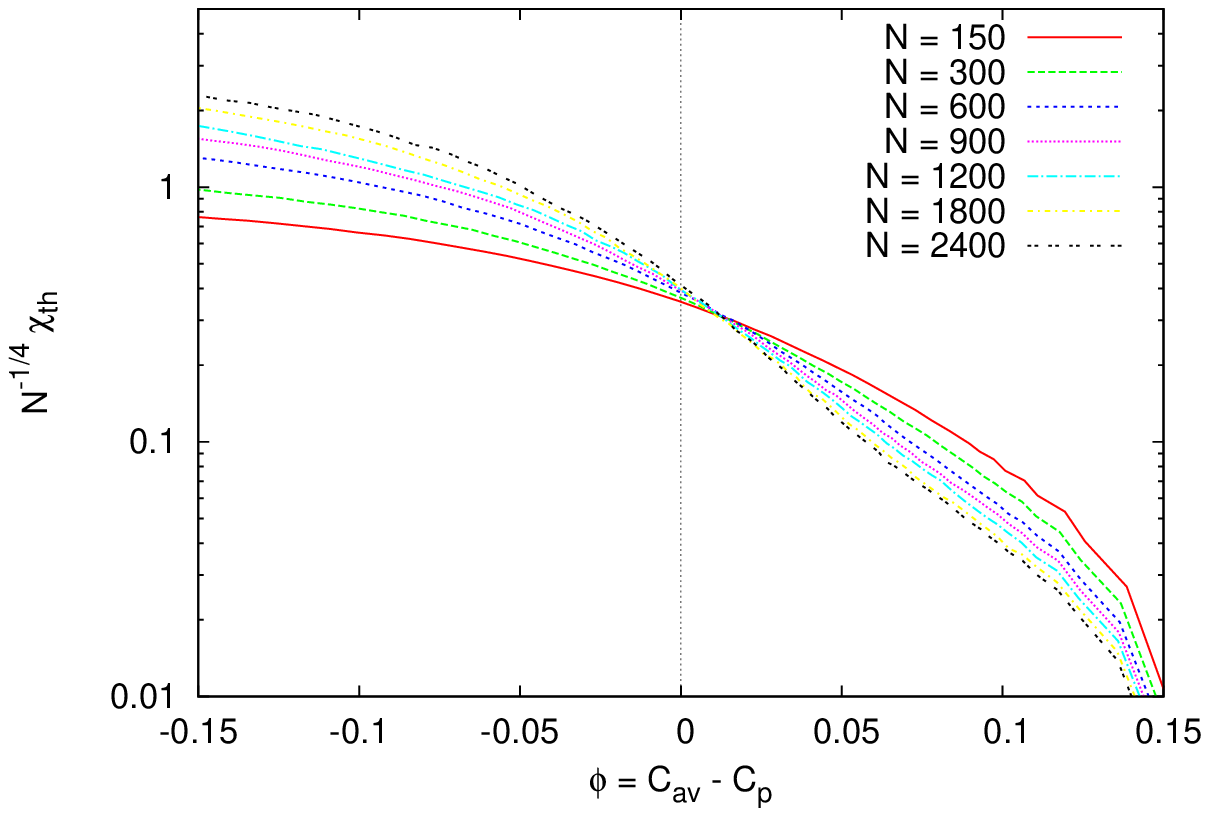} 
\caption{Rescaled heterogeneity susceptibility $\chi_{het}$ (left) and
  thermal susceptibility $\chi_{th}$ (right) in the gauged 3-spin
  model discussed in the text at the dynamical critical temperature
  $T_d$ as a function of average correlation $C_{av}$ for several
  sizes.  The crossing point moves towards $C_{av}=C_p$ (marked by a
  vertical line) for large $N$.}
\label{scaledGauged}
\end{center}
\end{figure}

These susceptibilities are shown in fig.~\ref{rawDataGauged}.  The
very large number of samples simulated allow to reduce the statistical
noise and to work on very clean data.  We notice that for large values of
$C_{av}$ the data converge, as expected, to a finite value in the
$N\to\infty$ limit: in this regime, dynamical fluctuations due to
heterogeneities are roughly one order of magnitude larger than those
due to thermal noise.

We now check finite size scaling of $\chi_{th}$ and $\chi_{het}$,
which for $\epsilon=0$ read
\begin{eqnarray}
\chi_{th}(\phi,0,N) &=& N^{1/4}\; h_{th}( \phi  N^{1/4})\;,\\
\chi_{het}(\phi,0,N) &=& N^{1/2}\; h_{het}( \phi  N^{1/4})\;. 
\end{eqnarray}
In figures \ref{scaledGauged} we plot the rescaled susceptibilities,
$\chi_{th} N^{-1/4}$ and $\chi_{het} N^{-1/2}$, as a function of
$\phi=C_{av}-C_p$ for various values of $N$.  We see that in agreement with
the analytical predictions the different curves cross very close to
$C=C_p=0.75$.  A more detailed analysis confirms that the crossing
point tends to $C_p$ for large $N$.

Before testing the $\phi N^{-1/4}$ scaling, we want to discuss the
main source of finite size effects in the gauged model. By choosing
the coupling signs independently, we have that energy fluctuations in
the initial configuration are $O(N^{-1/2})$, but with a rather large
coefficient if compared to the thermalized symmetric model.  The main
consequence is that, even for $N=O(10^3)$, the vast majority of
samples have either an initial energy much larger than $E_d\equiv -z/3
\tanh(1/T_d)$ (and thus decorrelate very fast), either much smaller
than $E_d$ (and thus are stuck at $C>C_p$). The final effect is to
have larger fluctuations and larger finite size effects, with respect
to a model where the initial energy is more concentrated around $E_d$.
Given that, in the thermodynamical limit, the energy must converge to
$E_d$, we introduce a new {\em fixed-energy} model where the initial
energy is fixed to $E_d$. 
This is achieved generating samples with a fixed number of
negative coupling equal to $M(1-\tanh 1/T_d)/2$. 

\begin{figure}[!ht]
\begin{center}
\includegraphics[width=0.6\textwidth]{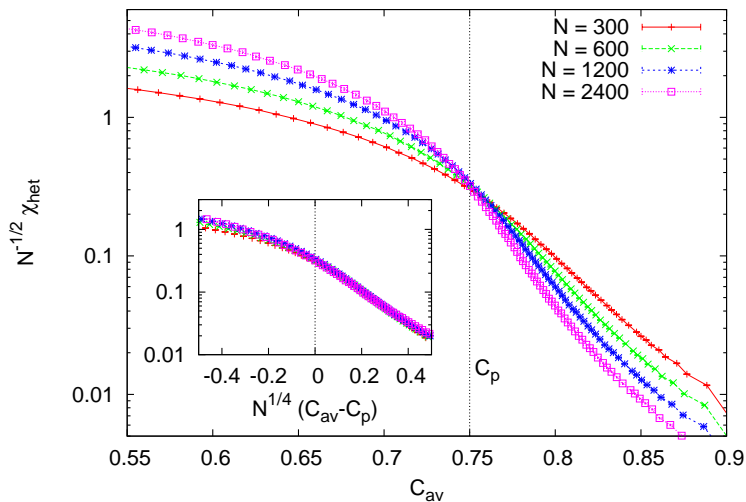}
\caption{Scaled heterogeneity susceptibility $\chi_{het} N^{-1/2}$ in the
  fixed-energy 3-spin model.}
\label{fig1}
\end{center}
\end{figure}

\begin{figure}[!ht]
\begin{center}
\includegraphics[width=0.6\textwidth]{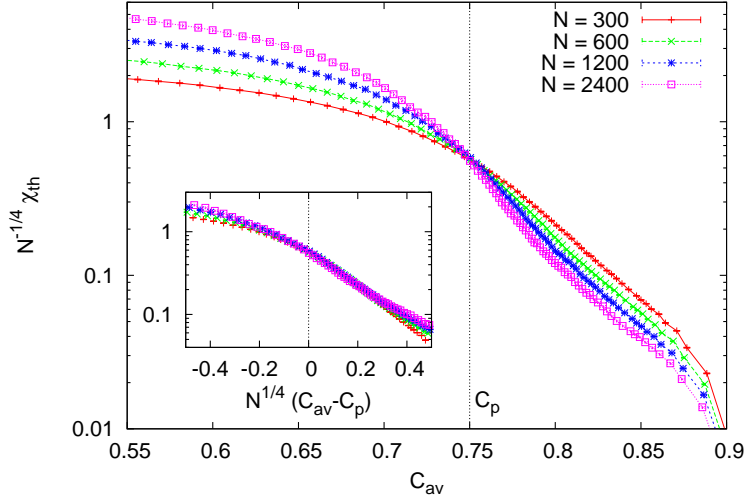}
\caption{Scaled thermal susceptibility $\chi_{het} N^{-1/4}$ in the
  fixed-energy 3-spin model.}
\label{fig2}
\end{center}
\end{figure}

We have simulated the fixed-energy model for sizes ranging from
$N=300$ to $N=2400$, with 2 thermal histories per sample and a number
of samples $N_S$ such that $N \times N_S = 3 \cdot 10^7$.  In the main
panels of Figs.~\ref{fig1} and \ref{fig2} we show the rescaled
susceptibilities $\chi_{th} N^{-1/4}$ and $\chi_{het} N^{-1/2}$ as a
function of the average correlation for several system sizes. A
comparison to the gauged model (see Fig.~\ref{scaledGauged}) shows
that finite size effects are reduced in this new model: indeed in
figures \ref{fig1} and \ref{fig2} data cross exactly at $C_{av}=C_p$
with almost no finite size corrections.

We test then the $\phi N^{-1/4}$ scaling in the insets of
Figs.~\ref{fig1} and \ref{fig2}, where the same rescaled
susceptibilities, are plotted now as a function of the scaling
variable $N^{-1/4}(C_{av}-C_p)$. We note that a good scaling is
observed in a relatively wide region around the origin.

\begin{figure}[!ht]
\begin{center}
\includegraphics[width= 0.6 \textwidth]{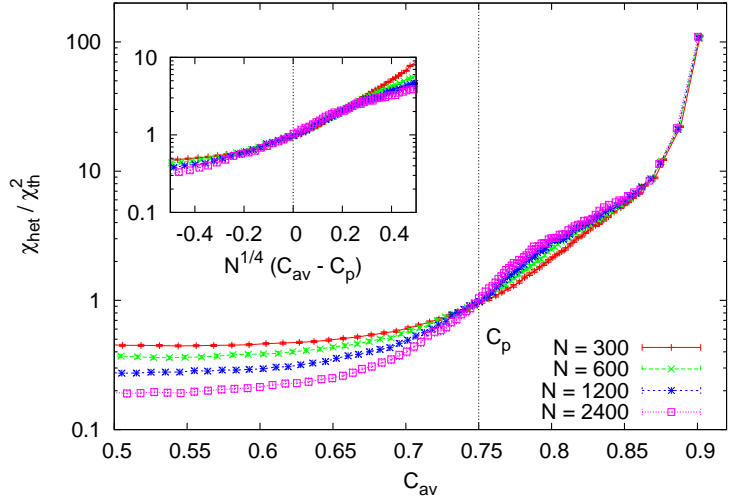}
\caption{Plot of the susceptibility ratio
  $\rho=\chi_{het}/\chi_{th}^2$ as a function of $C_{av}$. The curves
  cross in $C_p$, where, as seen in the inset, the slope scales as $N^{1/4}$. }
\label{chiRatio}
\end{center}
\end{figure}

In figure \ref{chiRatio} we show, as a function of $C_{av}$, the
$\rho$ ratio defined in Eq.(\ref{Rratio}), which, for $k=0$, is
\begin{equation}
\rho = \frac{\chi_{het}}{\chi_{th}^2}\;.
\end{equation}
As expected the curves cross at $C_p$. The limit $\lim_{C\to
  C_p}\lim_{N\to\infty} \rho(\phi,N)$ taken in the specified order, is
a direct measure of the variance of the random temperature entering
the field theory, $-(m_2+m_3)$. Unfortunately the values of $N$ we can
simulate do not allow an estimate of this limit.  However we can see
the scaling of slope $d\rho/dC$ in $C_p$ which behaves as $N^{1/4}$,
which confirms the formation of a discontinuity.

\subsection{Simulations in the alpha regime}

In order to study fluctuations in the alpha regime, the use of the
fixed-energy model is mandatory. Indeed the finite size effects in the
gauged model described above (strong fluctuations in the initial
configuration energy) amplify in the alpha regime, making the gauged
model almost useless. The fixed-energy model, on the contrary,
produces data showing a much better scaling. 

\begin{figure}[!ht]
\begin{center}
\includegraphics[width= 0.49 \textwidth]{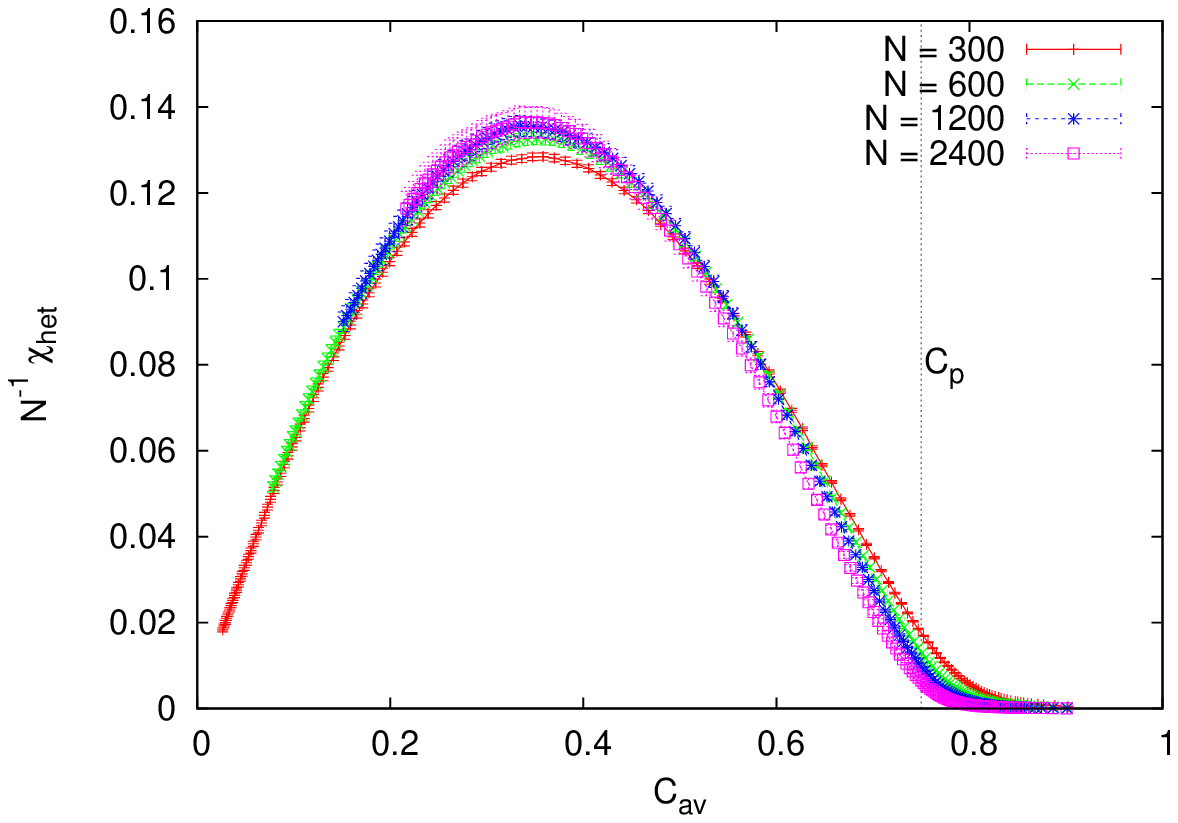}
\includegraphics[width= 0.49 \textwidth]{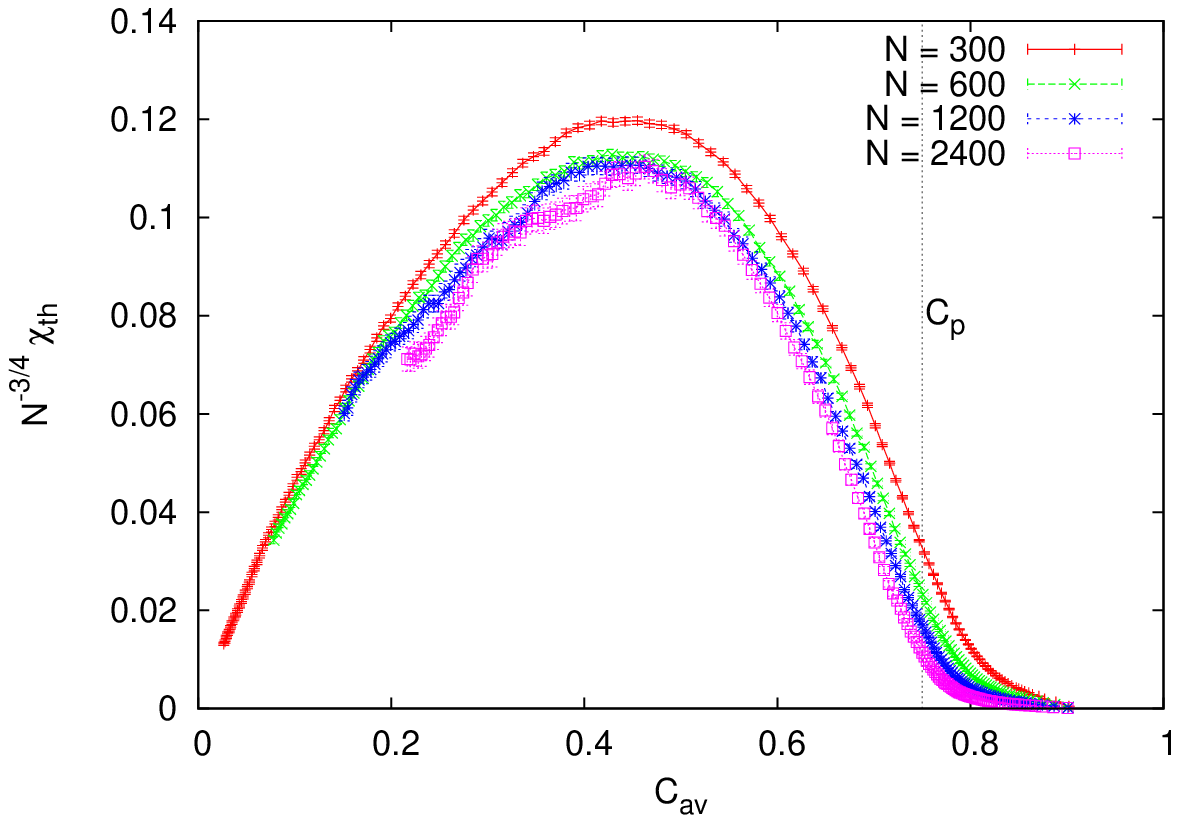}
\caption{Rescaled $N^{-1} \chi_{het}$ (left) and $N^{-3/4} \chi_{th}$
  (right) in the alpha regime.}
\label{scaledAlpha}
\end{center}
\end{figure}

We note first of all that, the different scaling between fluctuations
due to the heterogeneities and those due to the thermal noise persists
in the alpha regime. Indeed $\chi_{het}$ and $\chi_{th}$ show
different size dependence even for $C_{av} < C_p$, as can be seen in
Fig.~\ref{scaledAlpha}. While the scaling of $\chi_{het}$ is clearly
$O(N)$, that of $\chi_{th}$ has a smaller power: the data shown in
Fig.~\ref{scaledAlpha} (right) suggest the value $3/4$ argued for in
section~\ref{a-conj} to be an upper bound to the right power.

\begin{figure}[!ht]
\begin{center}
\includegraphics[width= 0.7 \textwidth]{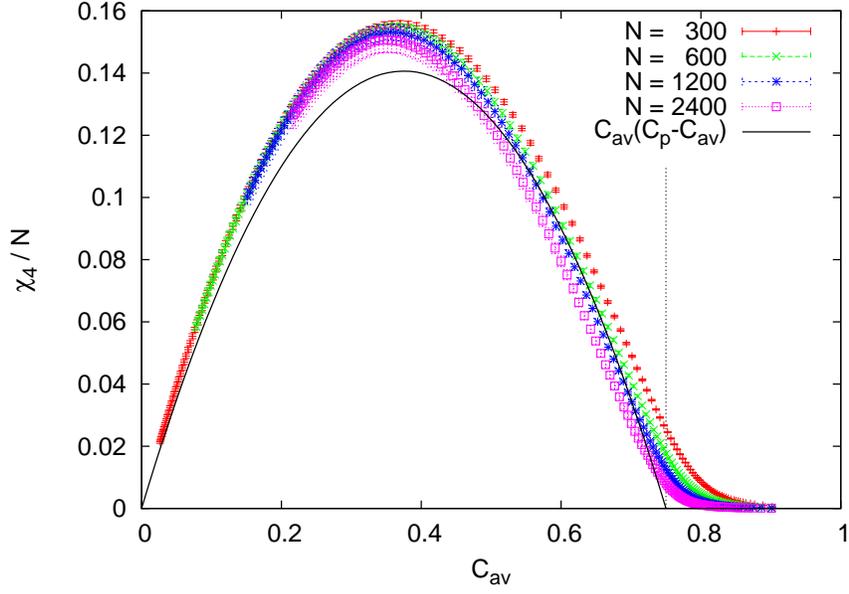}
\caption{Plot of the total rescaled susceptibility $\chi_4/N$ as a
  function of $C_{av}$ in the alpha regime. The curves for different
  values of $N$ are compared with the parabola $C_{av}(C_p-C_{av})$,
  derived under the hypothesis of a perfect bistability relaxation
  process.}
\label{chiTot}
\end{center}
\end{figure}

In Fig.~\ref{chiTot} we report data for the rescaled total
susceptibility $\chi_4 / N$ as a function of the average correlation.
The qualitative behaviour seems to reflect quite well the bistable
behaviour discussed in Section \ref{f-s-s}, and represented in
Fig.~\ref{chiTot} with the parabola $C_{av}(C_p-C_{av})$.  However
some deviations from this behavior are expected, especially for
$C_{av}$ close to $C_p$, because the time to relax from $C_p$ to 0 is
comparable to the time for entering the alpha regime in the fastest
samples: as explained, the final result should be that $\chi_4/N$ is
quadratic around $C_p$.

\begin{figure}[!ht]
\begin{center}
\includegraphics[width= 0.7 \textwidth]{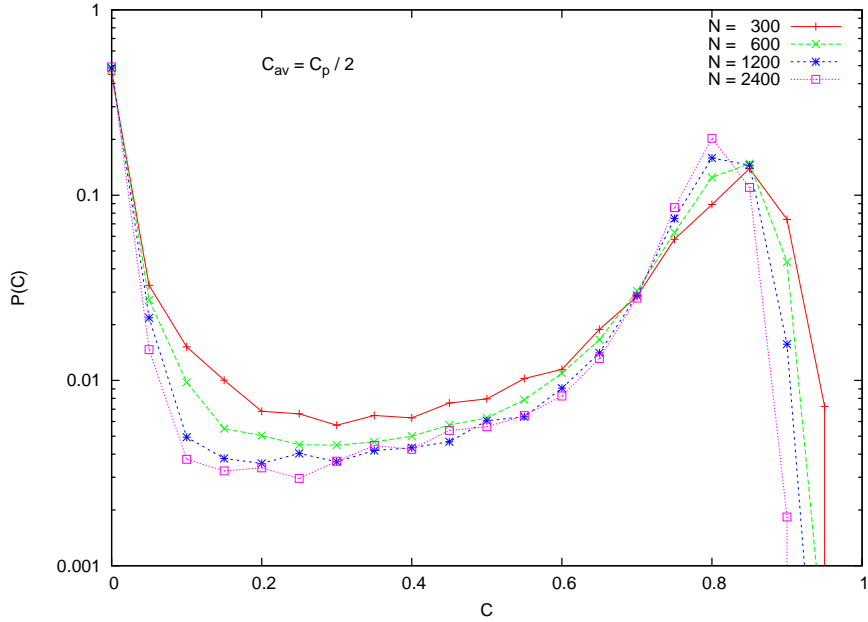}
\caption{Histogram of the correlation for $C_{av}=C_p/2$ on a
  logarithmic scale. While the data clearly show a bistable behavior,
  still a non vanishing part of the distribution between the two peaks
  seems to persists for large $N$. }
\label{histo}
\end{center}
\end{figure}

Bistability can be checked directly by looking at the histogram of
correlations. In fig.~\ref{histo} we report such histograms measured
at times such that the mean correlation satisfies $C_{av} =C_p/2$;
this time grows with system size. The shape of the histograms in
Fig.~\ref{histo} is clearly made of two well separated peaks.  However
one may notice that there is a small but non-zero probability
(apparently not vanishing when $N\to\infty$) of finding intermediate
values of correlation, and this is another plausible explanation for
the deviations of $\chi_4/N$ from the predicted parabola
$C_{av}(C_p-C_{av})$.

\section{Concluding remarks}
\label{concl}

In this paper we emphasize the importance of describing fluctuations
in a reparameterization invariant form in glassy systems.  We provide
a universal theory of these fluctuations in the beta regime close to
the mode coupling transition.

There are three main physical ingredients in our theory:
\begin{enumerate}
\item Time scale separation, leading to quasi-equilibrium sampling of
  metastable states.
\item The vicinity to a dynamical critical point.
\item Neglection of all possible non-perturbative effects.
\end{enumerate}

The first property allows us to study fluctuations through the use of
constrained equilibrium measures and their associated glassy effective
potential. These measures depend on a reference configuration which is
itself randomly chosen with canonical distribution. In this paper we
have extended the theory of the effective potential to study
fluctuation with respect to this source of noise.  We have considered
the effective potential for fixed initial configuration as a random
functional whose probability distribution relates to the one of
overlap fluctuations and therefore to dynamical heterogeneities.  In
our description time is eliminated and we use the average correlation
function as a clock.

The second property allows us to invoke universality and to use the 
general form of replica field theory that can be obtained by symmetry
considerations as an expansion around a Mode Coupling Transition
point.  The analysis of this theory leads to the identification of the
relevant fluctuation modes. The effective field theory that describes
them reduces, through tremendous simplifications to a scalar cubic
field theory with a local random field term. The random field term is
the expression of heterogeneity in the initial condition and acts as a
source of disorder that influences the subsequent dynamics. A
remarkable consequence of our description is that fluctuations with
respect to different sources of noise show different singular behavior
as the dynamic transition is approached.

The third point is quite delicate. In deriving the equivalence with
the RFIM all barrier jumping processes are neglected.  Quite naturally
one could hypothesize validity of the RFIM description beyond
perturbation theory. However, the dynamics in the non-perturbative
region, even its reparametrization invariant part, could be very
different in the RFIM and in supercooled liquids. The analogy between
liquid dynamics and the decay of metastable phases decay is only
partially valid. In ordinary first order transition kinetics,
competition between interface and volume free-energy leads to fast
growth of supercritical nuclei. This fast growth should not be present
in supercooled liquids. 

We tested our scenario in the favourable case of finite size scaling
in mean-field models. More work will be needed to test the scenario in
liquid models. In low dimension the critical -or pseudocritical-
properties of the spinodal point of RFIM can hardly be computed
analytically. One should therefore compare numerical results of
simulations of liquids with numerical results on the RFIM, however the
situation might be complicated by the fact that non-perturbative
effects could be different in the two kind of systems.

An important prediction of our theory is that the different components
of the dynamical fluctuations have different scaling
properties. Within a gaussian approximation we have found that the
heterogeneous susceptibility is proportional to the square of the
thermal one. It is not clear to us if this simple quadratic relation
holds beyond the gaussian approximation 
or two not-simply related exponents describe the corresponding singularities.  
This is an important point
that will need to be clarified through numerical simulations.
\vspace{.5 cm}

{\bf Acknowledgments} \\
\vspace{.3 cm}

We thank H. Castillo and G. Tarjus for discussions. SF acknowledges
the hospitality of the Dipartimento di Fisica, Sapienza Universit\`a
di Roma.

\end{document}